\documentclass{article}
\usepackage{amsmath, amssymb}
\usepackage{subcaption}
\usepackage{tikz,todonotes}
\usepackage[utf8]{inputenc}
\usepackage{hyperref}
\usepackage{cleveref}
\usepackage{algorithm}
\usepackage{algorithmic}
\usepackage{pgfplots}
\usepackage{frcursive}
\usepackage{amsmath,amsthm}
\usepackage{multirow}
\newtheorem{theorem}{Theorem}
 
\usepackage[margin=1in]{geometry}
\usepackage[sort&compress,square,comma,numbers]{natbib}

\DeclareFixedFont{\ttb}{T1}{txtt}{bx}{n}{12} 
\DeclareFixedFont{\ttm}{T1}{txtt}{m}{n}{12}  

\DeclareMathOperator{\E}{\mathbb{E}}
\DeclareMathOperator{\dbm}{\mathbf{\beta}_{\bsm}}

\DeclareMathOperator{\dkl}{D_\text{KL}}

\DeclareMathOperator{\bsxi}{\boldsymbol{\xi}}

\newcommand{\bE}{{\mathbb{E}}}

\newcommand{\bR}{{\mathbb{R}}}

\newcommand{\cA}{{\mathcal{A}}}

\newcommand{\cC}{{\mathcal{C}}}
\newcommand{\cD}{{\mathcal{D}}}
\newcommand{\cE}{{\mathcal{E}}}

\newcommand{\cK}{{\mathcal{K}}}
\newcommand{\cM}{{\mathcal{M}}}
\newcommand{\cN}{{\mathcal{N}}}

\newcommand{\cQ}{{\mathcal{Q}}}
\newcommand{\cV}{{\mathcal{V}}}

\newcommand{\bsm}{{\boldsymbol{m}}}
\newcommand{\bsu}{{\boldsymbol{u}}}

\newcommand{\bsw}{{\boldsymbol{w}}}

\newcommand{\bsy}{{\boldsymbol{y}}}

\newcommand{\bstheta}{{\boldsymbol{\theta}}}

\newcommand{\bsF}{{\boldsymbol{F}}}
\newcommand{\bsJ}{{\boldsymbol{J}}}

\newtheorem{example}{Example}

\usepackage{color}
\definecolor{deepblue}{rgb}{0,0,0.5}
\definecolor{deepred}{rgb}{0.6,0,0}
\definecolor{deepgreen}{rgb}{0,0.5,0}

\usepackage{listings}

\newcommand\pythonstyle{\lstset{
		language=Python,
		basicstyle=\ttm,
		otherkeywords={self},             
		keywordstyle=\ttb\color{deepblue},
		emph={MyClass,__init__},          
		emphstyle=\ttb\color{deepred},    
		stringstyle=\color{deepgreen},
		frame=tb,                         
		showstringspaces=false            %
}}

\lstnewenvironment{python}[1][]
{
	\pythonstyle
	\lstset{#1}
}
{}

\newcommand\pythoninline[1]{{\pythonstyle\lstinline!#1!}}

\title{Sequential infinite-dimensional Bayesian optimal experimental design with derivative-informed latent attention neural operator}
\author{Jinwoo Go, Peng Chen \footnote{ School of Computational Science and Engineering,
College of Computing,
Georgia Institute of Technology.
Address: 756 West Peachtree Street Northwest, Atlanta, GA 30308. \{jgo31, pchen402\}\@gatech.edu } }
\date{}

\begin{document}
  \maketitle

\begin{abstract}
   We develop a new computational framework to solve sequential Bayesian optimal experimental design (SBOED) problems constrained by large-scale partial differential equations with infinite-dimensional random parameters. We propose an adaptive terminal formulation of the optimality criteria for SBOED to achieve adaptive global optimality. We also establish an equivalent optimization formulation to achieve computational simplicity enabled by Laplace and low-rank approximations of the posterior. To accelerate the solution of the SBOED problem, we develop a derivative-informed latent attention neural operator (LANO), a new neural network surrogate model that leverages  (1) derivative-informed dimension reduction for latent encoding, (2) an attention mechanism to capture the dynamics in the latent space, (3) an efficient training in the latent space augmented by projected Jacobian, which collectively leads to an efficient, accurate, and scalable surrogate in computing not only the parameter-to-observable (PtO) maps but also their Jacobians. We further develop the formulation for the computation of the MAP points, the eigenpairs, and the sampling from posterior by LANO in the reduced spaces and use these computations to solve the SBOED problem. We demonstrate the superior accuracy of LANO compared to two other neural architectures and the high accuracy of LANO compared to the finite element method (FEM) for the computation of MAP points and eigenvalues in solving the SBOED problem with application to the experimental design of the time to take MRI images in monitoring tumor growth. We show that the proposed computational framework achieves an amortized $180\times$ speedup.
\end{abstract} 

\section{Introduction}

Bayesian optimal experimental design (BOED) is a powerful computational approach to optimally acquire information from experiments to understand complex systems under uncertainty through optimal design of experiments in a Bayesian framework. It is particularly prominent when the experiments are costly, time-consuming, or potentially dangerous. In these cases, we can only afford to conduct a limited number of experiments for data acquisition, e.g., in chemistry \cite{yonge2024model, walker2020chekipeuq, savara2020chekipeuq}, cognitive science \cite{myung2013tutorial}, clinical trials \cite{cheng2005bayesian, giovagnoli2021bayesian}, and engineering \cite{papadimitriou2004optimal}. BOED can be generally formulated as an optimization problem in optimizing some optimality criterion of the information gain or uncertainty of the system from the experimental or observational data \cite{atkinson2007optimum, ryan2016review, rainforth2024modern, huan2024optimal}. BOED maximizes the expected information gain (EIG) as an expectation of the Kullback--Leibler (KL) divergence between the posterior and prior distributions or minimizes the uncertainty of the system parameter measured by some statistics, e.g., trace or determinant of the posterior covariance, known as A-optimality or D-optimality. 

However, the solution of BOED problems faces significant computational challenges, especially for complex systems described by large-scale partial differential equation (PDE) models with high-/infinite-dimensional uncertain parameters. These challenges include but are not limited to (1) the optimality criteria, e.g., A-/D-/EIG optimalities, require the solution of a (possibly nonlinear) Bayesian inverse problem to compute the (possibly non-Gaussian) posterior distribution for each realization of the observation data; (2) each Bayesian inverse problem may involve numerous solutions of the PDE models for the evaluation of the parameter-to-observable (PtO) map at each step of the design optimization; (3) high-/infinite-dimensional BOED problems bring the curse of dimensionality, where the computational complexity may grow exponentially with respect to the dimensionality of the uncertain parameter in terms of the number of PDE solves; (4)
each PDE model may be costly to solve, which makes the solution of the BOED problem prohibitive; (5) the optimization of the experimental design is typically combinatorial and highly nonconvex, which becomes extremely difficult to solve for high-dimensional design variables. Many different computational
methods have been developed over the last decade in addressing these challenges, including (1) sparse polynomial chaos approximation of PtO maps
\cite{HuanMarzouk13,HuanMarzouk14},
(2) Laplace approximation of non-Gaussian posterior distributions
\cite{AlexanderianPetraStadlerEtAl16,
BeckDiaEspathEtAl18, LongScavinoTemponeEtAl13a, LongMotamedTempone15,
BeckMansourDiaEspathEtAl20}, (3) low-rank approximation of prior-preconditioned Hessian of the data misfit term \cite{AlexanderianPetraStadlerEtAl14,AlexanderianGloorGhattas16,AlexanderianPetraStadlerEtAl16,SaibabaAlexanderianIpsen17,CrestelAlexanderianStadlerEtAl17,AttiaAlexanderianSaibaba18}, 
(4) reduced order models
\cite{Aretz-NellesenChenGreplEtAl20, AretzChenVeroy21,AretzChenDegenEtAl24} 
and deep neural networks \cite{wu2023large,go2023accelerating} that serve as surrogate models of the PDEs or PtO maps, (5) variational inference and neural estimation for fast approximation of the EIG or mutual information \cite{FosterJankowiakBinghamEtAl19,
KleinegesseGutmann20,GoIsaac22,ShenHuan2023,orozco2024probabilistic}, and (6) efficient optimization methods using gradients  \cite{AlexanderianPetraStadlerEtAl14,AlexanderianPetraStadlerEtAl16,HuanMarzouk14}, greedy \cite{JagalurMohanMarzouk21,helin2022edge,AretzChenVeroy21,AretzChenDegenEtAl24} and swapping greedy algorithms \cite{WuChenGhattas23,wu2023fast,wu2023large}, and their combination \cite{go2023accelerating}.

Despite these advancements, it remains a critical challenge and an open research area for most of the computational methods mentioned above to solve BOED problems sequentially, where experiments are designed and conducted adaptively based on previous outcomes for complex dynamical systems. There have been two main approaches in formulating sequential BOED (SBOED) problems: static approach and adaptive approach \cite{RolandGautier2000, tao2003adaptive}. The static approach considers all possible experimental outcomes upfront, designing the entire sequence of experiments before any experiments are conducted. 
In contrast, the adaptive approach designs each experiment sequentially, updating model parameters after each observation before designing the next experiment. It can be formulated as to optimize one step ahead (myopic, greedy) 
\cite{drovandi2014sequential, kim2014hierarchical, vincent2017darc, surer2024sequential, murphy2003optimal} or multiple steps ahead using back induction (dynamical programming) 
\cite{ huan2016sequential,foster2021deep,ivanova2021implicit, shen2023variational}. 
One recent promising approach to solving the SBOED problem is the use of reinforcement learning \cite{foster2021deep,ivanova2021implicit,blau2022optimizing, shen2023variational}. However, the high computational cost of solving PDEs and the curse of dimensionality make a direct application of these methods infeasible to SBOED problems constrained by large-scale PDEs with high-dimensional  parameters. 

To address these combined challenges of SBOED problems, we propose using a surrogate-based approach, particularly based on neural operators \cite{kovachki2023neural}. Neural operators are deep learning models designed to learn a mapping between function spaces, making them suited for solving PDEs and related tasks in high-dimensional settings \cite{li2020fourier, lu2019deeponet,o2024derivative}. 
While neural operators address function space mapping, attention models \cite{vaswani2017attention} have shown strong performance in handling sequential data and capturing long-range dependencies, as demonstrated in language models like GPT \cite{achiam2023gpt} and Llama \cite{touvron2023llama}. Recognizing the potential of combining these approaches, several advanced methods applying attention mechanisms to neural operators have been proposed. Examples include OFormer \cite{li2022transformer}, GNOT \cite{hao2023gnot}, and ViTo \cite{ovadia2024vito}, which offer improved handling of complex, multi-scale problems while maintaining high accuracy. These hybrid approaches could potentially enhance the efficiency and effectiveness of SBOED for PDE systems by better capturing temporal dependencies and multi-scale interactions in the underlying physical processes. In addition to the attention model, latent dynamics approaches provide another avenue for efficiency to reduce the dimension of neural networks and increase accuracy. Stemming from neural ODEs \cite{chen2018neural} and other latent models \cite{regazzoni2023latent}, these approaches provide alternatives to traditional ResNet architectures \cite{he2016deep} or direct parameters to all-time step observable mappings \cite{li2020fourier}. 
Moreover, DIPNet \cite{o2022derivative} incorporates derivative information in dimension reduction, enhanced by further incorporating derivative information for neural network training in DINO \cite{o2024derivative} and DE-DeepONet \cite{qiu2024derivative}, which have been demonstrated to improve the accuracy of not only the output but also its derivative and have been applied in solving inverse, optimization, and BOED problems \cite{cao2024efficient, luo2023efficient, go2023accelerating}. 
These collective developments in surrogate modeling offer promising directions in addressing the challenges in solving high-/infinite-dimensional SBOED problems constrained by large-scale PDE models.

\textbf{Contributions}: To solve the infinite-dimensional SBOED constrained by large-scale dynamical systems described by PDEs, we develop a fast, scalable, and accurate computational framework with the following contributions: (1) we propose a new adaptive terminal formulation of SBOED problem to calculate globally optimal design conditioned on a stream of observed data at every adaptive step; (2) we establish an equivalent formulation of the adaptive SBOED problem in terms of conditional EIG measured by the KL divergence between the  posterior and the prior distributions, which significantly simplifies the evaluation of the optimality criteria at every optimization step; (3) we formulate a scalable approximation framework to solve the adaptive SBOED problem with infinite-dimensional parameters by Laplace approximation of the posterior, a low-rank approximation of the posterior covariance, and an adaptive optimization algorithm to minimize the conditional EIG; (4) we develop a novel surrogate model named latent attention neural operator (LANO) that leverages latent encoding of the high-dimensional input parameter and output observable by derivative-informed dimension reduction, latent attention mechanism in capturing the temporal correlation of the latent variables, and latent dynamics based on the attention architecture in approximating both the PtO maps and their Jacobians; (5) we derive LANO-enabled efficient computation of the optimality criteria, including computing the MAP point, solving the eigenvalue problem, and sampling from the adaptive posterior, all in reduced spaces with small input and output dimensions; (6) we present numerical experiments for the demonstration of the accuracy and efficiency of our proposed computational framework in solving the SBOED problem, with an application in optimally conducting MRI experiments to monitor tumor growth. In particular, we report the comparison of LANO with DIPNet and neural ODE and show the much more accurate approximation of the PtO map and its Jacobian by LANO. We demonstrate the high accuracy of the LANO-enabled computation of the MAP point and eigenvalues compared to a high-fidelity computation using a FEM. We apply our proposed method to solve the SBOED problem and demonstrate its effectiveness in reducing the uncertainty of the parameters compared to an intuitive experimental design. For this example, we demonstrate its efficiency in achieving an amortized $180\times$ computational speedup, accounting for both online evaluation time and offline time in data generation and training. 

The following part of the paper is organized as follows. In Section \ref{sec:problem}, we present the formulation of SBOED with a new adaptive terminal formulation, followed by Section \ref{sec:scalableSBOED} to present the SBOED problem. We introduce a novel LANO surrogate in efficient computation of the optimality criteria in Section \ref{sec:LANO}. We demonstrate the accuracy and efficiency of the proposed method for solving an application problem of designing MRI experiments to monitor tumor growth in Section \ref{sec:numerics} and conclude in Section \ref{sec:conclusion}.

\section{Problem formulation}
\label{sec:problem}
This section introduces infinite-dimensional Bayesian inverse problems constrained by dynamical systems represented as time-dependent PDEs, where the uncertain parameter is a random field. We then present different formulations of the SBOED problem for optimal data acquisition to minimize the uncertainty of the model parameter in the context of Bayesian inverse problems.
\subsection{Bayesian inverse problem}

We consider Bayesian inverse problems governed by time-dependent PDEs with infinite-dimensional uncertain parameters, which can be generally written as 
\begin{equation}\label{eq:pde}
    \partial_tu(t,x) + R(u(t,x),m(x)) = 0, \quad (t,x) \in (0, T] \times \Omega,
\end{equation}
where $T > 0$ is a terminal time, $\Omega \subset \mathbb{R}^{d_x}$ is an open bounded physical domain in dimension $d_x$, $u(t) \in V$ is the state variable in Hilbert space $V$ defined in $\Omega$ with proper boundary condition for every time $t\in (0, T)$, $u(0) = u_0$ is an initial condition at time $t = 0$, $m \in M$ is a random field parameter in Hilbert space $M$, $R:V \times M \rightarrow V'$ is a differential operator, where $V'$ is the dual space of $V$. 

We introduce a partition of the time interval $[0, T]$ into $K$ sub-intervals $[t_{k-1}, t_k]$, $k = 1, \dots, K$, with
$0=t_0<t_1<...<t_K=T$. Then, we can define the discrete-time state variable as
$u_k(x) = u(t_k, x)$  for $k = 0, ...,K$, and a corresponding $d_y$-dimensional parameter-to-observable (PtO) map at time $t_k$ as $\mathcal{F}_k:M \rightarrow \mathbb{R}^{d_y}$ for $k = 1, ..., K$, typically given as $\mathcal{F}_k(m) = \mathcal{B}_k (u_k(m))$, where $\mathcal{B}_k : V \to \mathbb{R}^{d_y}$ is an observation operator, $u_k$ is the solution of the PDE \eqref{eq:pde} at time $t_k$ and parameter realization $m$.

We consider noisy observation data $\bsy_k$ at time $t_k$ corrupted by additive noise as 
\begin{equation}\label{eq:obsdata}
    \bsy_k = \mathcal{F}_k(m) + \boldsymbol{\epsilon}_k, \quad \text{for } k=1, \dots, K,
\end{equation}
where we assume that the observation noise $\boldsymbol{\epsilon}_k$ follows a Gaussian distribution $\cN (\mathbf{0},\Gamma_\text{noise})$ with covariance matrix $\Gamma_\text{noise} \in \mathbb{R}^{d_y\times d_y}$. Under this assumption, the likelihood function of the data $\bsy = (\bsy_1, \ldots, \bsy_K)$ reads
\begin{equation}
    \pi_\text{like}(\bsy|m) = \frac{1}{\sqrt{(2\pi)^K|\Gamma_\text{noise}|)}}\exp{(-\Phi(\bsy,m))}, 
\end{equation}
where $|\Gamma_\text{noise}|$ is the determinant of the noise covariance, and $\Phi(\bsy, m)$ is a potential function representing the misfit between the observation data and the parameter-to-observable map, given by
\begin{equation}
    \Phi(\bsy,m) = \frac{1}{2}\sum_{k=1}^K||\bsy_k - \mathcal{F}_k(m)||_{\Gamma_\text{noise}^{-1}}^2,
\end{equation}
where $\|v\|_{\Gamma_\text{noise}^{-1}}^2 = v^T \Gamma_\text{noise}^{-1} v$ for any vector $v \in \mathbb{R}^{d_y}$.

For the random field model parameter $m$, we consider a Gaussian prior $\mu_\text{prior} = \cN(m_\text{prior}$, $ \mathcal{C}_\text{prior})$ with mean $m_\text{prior}$ and a Mat\'ern covariance operator $\mathcal{C}_\text{prior} = \mathcal{A}^{-\alpha}$, where $\mathcal{A} = -\gamma\Delta+\delta I$ is defined on $\Omega$ with a proper (e.g., Robin) boundary condition \cite{daon2016mitigating}, $\alpha > d_x/2$ such that the covariance operator $\mathcal{C}_\text{prior}$ is of trace class. Here, $\alpha, \gamma, \delta > 0$ are the parameters that collectively determine the smoothness, variance, and correlation length of the random field.

The posterior measure $\mu_\text{post}$ of the parameter $m$ conditioned on the observation data $\bsy$ is given by Bayes' rule using the Radon--Nikodym derivative as
\begin{equation}
 \frac{d\mu_\text{post}}{ d\mu_\text{prior}} = \frac{1}{\pi(\bsy)}\pi_\text{like}(\bsy|m),   
\end{equation}
where $\pi(\bsy)$ is the marginal likelihood (or evidence) given by the infinite-dimensional integral of the likelihood function over the prior distribution, i.e.,
\begin{equation}
    \pi(\bsy) = \int_{M}\pi_\text{like}(\bsy|m)d\mu_\text{prior}(m),
\end{equation}
which is typically intractable to compute due to the high/infinite dimensionality of the parameter space $M$. The central task of the Bayesian inverse problems is to draw samples from the posterior distribution $\mu_\text{post}$ to quantify the uncertainty of the model parameter $m$ and its related quantity of interest.

\subsection{Sequential Bayesian optimal experimental design}

We consider sequential experimental design in the context of Bayesian inverse problems, where the goal is to design the optimal experiment $\bsxi^*$ to acquire the most informative data that minimizes the uncertainty of the parameter or maximizes the information about the parameter gained from the data. For simplicity, we consider the design problem of selecting the $d < K$ most informative time steps out of the $K$ time steps to make observations, with the design space $\Xi$ defined as 
\begin{equation}\label{eq:design-space}
    \Xi := \left\{\bsxi = (\xi_1, \dots, \xi_K) \in \{0, 1\}^{K}: \ \sum_{k=1}^{K}\xi_k = d\right\},
\end{equation}
where $\xi_k = 1$ represents that we select the $k$-th time step to make observation, or use the data $\bsy_k$, and $\xi_k = 0$ otherwise. In a more general setting, we can also consider the design problem of selecting both the observation time steps and observation locations simultaneously. Under the experimental design $\bsxi$, we denote the prior, the posterior, the likelihood, and the marginal likelihood for simplicity as $\mu(m)$, $\mu(m|\bsy, \bsxi)$, $\pi(\bsy|m,\bsxi)$, and $\pi(\bsy|\bsxi)$, respectively. 

In the so-called static SBOED \cite{foster2019variational, foster2021deep,kleinegesse2020bayesian, wu2023fast, go2023accelerating}, the goal is to find the optimal experimental design $\bsxi^*$ that maximizes the expected information gain about the model parameter $m$ in one step, i.e.,
\begin{equation}\label{eq:staticBOED}
    \bsxi^* = \arg\max_{\bsxi\in \Xi} \E_{\pi(\bsy|\bsxi)} [I(\bsxi)],
\end{equation}
where $I(\bsxi)$ represents an information gain defined as the Kullback-Leibler divergence 
\begin{equation}
    I(\bsxi) := \dkl(\mu(m|\bsy, \bsxi)||\mu(m)) =  \int_{M} \log\left(\frac{d\mu(m|\bsy,\bsxi)}{d\mu(m)}\right) \mu(d m|\bsy,\bsxi),
\end{equation}
which measures the information gain from the prior measure $\mu(m)$ to the posterior measure $\mu(m|\bsy,\bsxi)$. This static formulation of SBOED does not consider the sequential and time-dependent nature of the experimental design, and it is not adaptive to the information gained from previous observations. 
This static formulation of SBOED is not adaptive to the information gained from previous observations. 

In contrast to the static formulation of SBOED, the design of the experiment is adaptive and conditioned on the data from all previous observations in a sequential formulation of SBOED. Let $\bsy_{1:i} = (\bsy_1, \dots, \bsy_i)$ and $\bsxi_{1:i} = (\xi_1, \dots, \xi_i)$ denote the data and the experimental design up to time $t_i$, respectively. We can define the incremental information gain for the experimental design $\xi_i$ at time $t_i$ as 
\begin{equation}\label{eq:incre_form}
    I(\xi_i) = \dkl(\mu(m|\bsy_{1:i}, \bsxi_{1:i})||\mu(m|\bsy_{1:i-1}, \bsxi_{1:i-1})), \quad i = 2, \dots, K,
\end{equation}
and $I(\xi_1) = \dkl(\mu(m|\bsy_{1}, \bsxi_{1})||\mu(m))$. 
An incremental formulation of SBOED aims to maximize the expected incremental information gain at each time step $t_i$, i.e.,
\begin{equation}\label{eq:greedyBOED}
    \xi_i^* = \arg\max_{\xi_i} \E_{\pi(\bsy_i|\xi_i)} [I(\xi_i)], \quad \text{for } i = 1, \ldots, K,
\end{equation}
where the expectation is taken with respect to the marginal likelihood of the data $\bsy_i$ given design $\xi_i$, i.e., $\pi(\bsy_i|\xi_i) = \int_{\cM} \pi(\bsy_i|m, \xi_i) \mu(dm|\bsy^*_{1:i-1}, \bsxi^*_{1:i-1})$, with the data $\bsy^*_{1:i-1}$ observed at the optimized design $\bsxi^*_{1:i-1}$. 

This greedy algorithm is adaptive and responsive to the information gained from each experiment, allowing for flexibility in experimental planning. However, this approach may not yield the optimal solution regarding the total expected information gain across all experiments, as it does not consider the cumulative effect of its choices. Meanwhile, this incremental formulation is more challenging to justify and implement when selecting the time to make observations, as in our application, than when selecting the most informative spatial locations or sensors to make observations at each predefined time step. 

\subsection{Adaptive terminal formulation of SBOED}
To achieve the adaptive global optimality of the sequential experimental design, we solve an adaptive SBOED problem, as in the following example, to select $d = 4$ out of $K = 10$ observation times. 
\begin{example}\label{ex:adaptiveSBOED}
    We first solve the static SBOED to get $\bsxi^*$. Then we move to the time at which we have the first nonzero entry of $\bsxi^*$, e.g., with $\bsxi^* = (0, 0, 1, 0, 0, 1, 1, 0, 1, 0)$, we move to time $t_3$. Then we make observation $\bsy_3^*$ at $t_3$ and solve the next SBOED problem to select 3 out of 7 observation times from time $t_4$ and on. This is done by minimizing an expected cumulative information gain for the rest of the 3 experiments to be designed from $t_4$. We repeat the adaptive optimization process until all the observations are made.
\end{example}

Let $t_{i-1}$ denote the time that the last observation is made. Let $\bsxi_{1:i:K} = (\xi_1^*, \dots, \xi_{i-1}^*$, $\xi_i, \dots, \xi_K)$ denote the experimental design at all time steps, with optimized design $\bsxi_{1:i-1}^* = (\xi_1^*, \dots, \xi_{i-1}^*)$ before time $t_i$ and the design $\bsxi_{i:K} = (\xi_i, \dots, \xi_K)$ to be optimized from $t_i$ to $t_K$. Let $\bsy_{1:i:K} = (\bsy_1^*, \dots, \bsy_{i-1}^*,  \bsy_i, \dots, \bsy_K)$ denote the observation data corresponding to the design $\bsxi_{1:i:K}$, with $\bsy_{1:i-1}^* = (\bsy_1^*, \dots, \bsy_{i-1}^*)$. Note that the data $\bsy_k^*$, $k = 1, \dots, i-1$, are observed only when $\xi_k^*$ is not zero. We use $\bsy_{1:i-1}^*$ for notational convenience. Then, the SBOED based on the expected cumulative information gain can be formulated as
\begin{equation}\label{eq:cumulativeBOED}
    \bsxi_{i:K}^* = \arg\max_{\bsxi_{i:K}} \E_{\pi(\bsy_{i:K}|\bsxi_{1:i:K}, \bsy_{1:i-1}^*)} 
    \left[
        \sum_{k=i}^K I(\xi_k)
    \right], 
\end{equation}
where the expectation is taken with respect to the marginal likelihood of the data $\bsy_{i:K}$ for the design $\bsxi_{i:K}$. This approach seeks to find an optimal trajectory of experimental setups that maximizes the cumulative information gain. However, a direct solution to this optimization problem may be prohibitive due to the cumulative computation of the information gain. We establish the following equivalent optimization problem with a terminal formulation of the objective function, which is much simpler to compute than the cumulative formulation. See the proof in Appendix \ref{sec:equivalence}. 
\begin{theorem}\label{thm:equivalence}
    Let $\mu(m|\bsy_{1:i:K}, \bsxi_{1:i:K})$ denote the posterior distribution for the observations $\bsy_{1:i:K}$ given experimental design $\bsxi_{1:i:K}$, then the optimization problem \eqref{eq:cumulativeBOED} is equivalent to the following optimization problem     \begin{equation}\label{eq:terminalinitialBOED}
    \bsxi_{i:K}^* = \arg\max_{\bsxi_{i:K}} \E_{\pi(\bsy_{i:K}|\bsxi_{1:i:K},\bsy_{1:i-1}^*)} 
    \left[
    \dkl(\mu(m|\bsy_{1:i:K}, \bsxi_{1:i:K})||\mu(m))\right].
    \end{equation}
\end{theorem}

\section{Scalable approximations for SBOED}\label{sec:scalableSBOED}

In this section, we present scalable approximation methods to solve the SBOED, including high-fidelity discretization of the random field parameter, Laplace approximation of the posterior distribution, low-rank approximation of the posterior covariance, as well as the resulting approximation of the optimality criteria introduced in the last section for the sequential experimental design.

\subsection{High-fidelity discretization}
To solve the infinite-dimensional inverse problem, we introduce a high-fidelity discretization using FEM to approximate the random field parameter $m$ in a finite-dimensional subspace $M_{d_m} \subset M$ of dimension $d_m$ \cite{go2023accelerating, wu2023fast}. This space is spanned by piecewise continuous Lagrange polynomial basis functions $\{\phi_i\}_{k=1}^{d_m}$. The basis is defined over a mesh of the domain $\Omega$ at vertices $\{x_j\}_{j=1}^{d_m}$, such that $\phi_i(x_j) = \delta_{ij}$ and $i,j = 1,\ldots,d_m$. The approximation of the model parameter $m \in M$ in $M_{d_m}$, denoted as $\hat{m}$, can be expressed as
\begin{equation}
\hat{m}(x) = \sum_{k=1}^{d_m} m_i \phi_i(x), \quad x \in \Omega.
\end{equation}
We denote $\bsm = (m_1,\dots,m_{d_m})^T \in \mathbb{R}^{d_m}$ as the coefficient vector, and denote $F_k:\mathbb{R}^{d_m} \rightarrow \mathbb{R}^{d_y}$ as the discrete version of the PtO map $\mathcal{F}_k$ correspondingly. 
Moreover, we denote $\mathbb{M} \in \mathbb{R}^{d_m \times d_m}$ and $\mathbb{A} \in \mathbb{R}^{d_m \times d_m}$ as the finite element mass matrix and stiffness matrix given by 
\begin{equation}\label{eq:mass}
  \mathbb{M}_{ij} = \int_D \phi_i(x) \phi_j(x) dx, \quad i,j = 1,\ldots,d_m,
\end{equation}
and
\begin{displaymath}
    \mathbb{A}_{ij} = \int_D (\gamma \nabla \phi_i(x) \cdot \nabla \phi_j(x) + \delta \phi_i(x) \phi_j(x)) dx, \quad i,j = 1,\ldots,d_m.
\end{displaymath}
Then the discrete parameter $\bsm$ follows a Gaussian prior distribution $\cN(\bsm_\text{prior}, \Gamma_\text{prior})$ with $\bsm_\text{prior}$, discretized form of $m_\text{prior}$, and the covariance matrix given by $\Gamma_\text{prior} = \mathbb{A}^{-1}\mathbb{M}\mathbb{A}^{-1}$ \cite{Bui-ThanhGhattasMartinEtAl13}. We also use finite element spatial discretization to approximate the state variable $u$ in the PDE \eqref{eq:pde} and the corresponding observation operator $F_k$.

\subsection{Laplace approximation of the posterior distribution}
\label{sec:Laplace}
We consider a Laplace approximation of the posterior distribution of the discrete parameter $\bsm$ conditioned on a general observation data $\bsy = (\bsy_1, \dots, \bsy_K)$ for a given experimental design $\bsxi = (\xi_1, \dots, \xi_K)$, which is denoted as $\pi(\bsm|\bsy, \bsxi) = \cN(\bsm_\text{MAP}^{\bsy,\bsxi}, \Gamma_\text{post}^{\bsy,\bsxi})$, where the maximum-a-posteriori (MAP) point $\bsm_\text{MAP}^{\bsy,\bsxi}$ is given as the solution of the optimization problem
\begin{equation}\label{eq:MAP}
\bsm_{\text{MAP}}^{\bsy,\bsxi} := \arg\min_{\bsm} \frac{1}{2} \sum_{k=1}^K \xi_k ||\bsy_k-F_k(\bsm)||_{\Gamma_\text{noise}^{-1}}^2 + \frac{1}{2}||\bsm-\bsm_\text{prior}||_{\Gamma_\text{prior}^{-1}}^2,
\end{equation}
e.g., using an inexact Newton-CG algorithm \cite{villa2021hippylib}, which is scalable with respect to the dimension of the parameter $\bsm$, and the covariance matrix $\Gamma_\text{post}^{\bsy,\bsxi}$ is given by
\begin{equation}\label{eq:postcov}
    \Gamma_\text{post}^{\bsy,\bsxi} = (H_\text{misfit}^{\bsy,\bsxi}(\bsm_\text{MAP}^{\bsy,\bsxi}) + \Gamma_\text{prior}^{-1})^{-1},
\end{equation}
where $H_\text{misfit}^{\bsy,\bsxi}$ is the Hessian of the misfit term evaluated at $\bsm = \bsm_\text{MAP}^{\bsy,\bsxi}$. In practice, we often consider a Gauss--Newton (GN) approximation of the Hessian $H_\text{misfit}^{\bsy,\bsxi}(\bsm_\text{MAP}^{\bsy,\bsxi})$ as \cite{wu2023fast}
\begin{equation}\label{eq:GN}
    H^{\text{GN}, \bsxi}_\text{misfit}(\bsm_\text{MAP}^{\bsy,\bsxi}) = \sum_{k=1}^K \xi_k \nabla_{\bsm}F_k(\bsm_\text{MAP}^{\bsy,\bsxi})^T\Gamma_\text{noise}^{-1}\nabla_{\bsm}F_k(\bsm_\text{MAP}^{\bsy,\bsxi}),
\end{equation}
with $\nabla_{\bsm}F_k(\bsm_\text{MAP}^{\bsy,\bsxi})$ denoting the Jacobian of the observable $F_k$ evaluated at $\bsm = \bsm_\text{MAP}^{\bsy,\bsxi}$. Note that the above MAP point and posterior covariance matrix are defined for the data and experimental design across all the time steps. Up to time $t_i$, with the observed data $\bsy_{1:i-1}^*$ for the optimized experimental design $\bsxi_{1:i-1}^*$, we denote the MAP point and the posterior covariance matrix as $\bsm_{\text{MAP}}^{(i-1)}$ and $\Gamma_{\text{post}}^{(i-1)}$, with the sum from $k = 1$ to $K$ in \eqref{eq:MAP} and \eqref{eq:GN} replaced by that from $k = 1$ to $i-1$, respectively.

\subsection{Low-rank approximation of the posterior distribution}
\label{sec:low-rank}
To compute the large posterior covariance matrix $\Gamma_{\text{post}} \in \mathbb{R}^{d_m \times d_m}$ with a high dimension $d_m$, we employ a low-rank approximation of the Hessian misfit $H^{\text{GN}, \bsxi}_{\text{misfit}}$ in \eqref{eq:GN} by solving a generalized eigenvalue problem as
\begin{equation}\label{eq:geneig}
    H^{\text{GN}, \bsxi}_{\text{misfit}}(\bsm_\text{MAP}^{\bsy,\bsxi}) \bsw_j = \lambda_j \Gamma_{\text{prior}}^{-1} \bsw_j, \quad j = 1, \ldots, r,    
\end{equation}
using, e.g., a randomized algorithm \cite{villa2021hippylib}, which is scalable with respect to $d_m$. Here the eigenvalues $\lambda_1 \geq \cdots \geq \lambda_r > 0$ with $r$ such that $\lambda_r \ll 1$, and the corresponding eigenvectors satisfy $\bsw_i^T \Gamma_{\text{prior}}^{-1} \bsw_j = \delta_{ij}$. To this end, the posterior covariance matrix $\Gamma_{\text{post}}$ in \eqref{eq:postcov} can be approximated as \cite{go2023accelerating, villa2021hippylib}

\begin{equation}\label{eq:lowrankGamma}
    \Gamma_{\text{post}}^{\bsy,\bsxi} \approx \Gamma_{\text{prior}} - W_r D_r W_r^T,
\end{equation}
where $W_r = [\bsw_1, \ldots, \bsw_r]$, 
$D_r = \text{diag}(d_1, \dots, d_r)$ with $d_j = \lambda_j / (\lambda_j + 1)$, $j = 1, \dots, r$.
Similarly, up to before time $t_i$, we denote these quantities as $W_r^{(i-1)}$, 
and $D_r^{(i-1)}$, 
corresponding to the posterior covariance matrix $\Gamma_{\text{post}}^{(i-1)}$ as in the last section. With this low-rank approximation, we can draw a random sample from the Laplace approximation of the posterior distribution $\cN(\bsm_\text{MAP}^{\bsy,\bsxi}, \Gamma_\text{post}^{\bsy,\bsxi})$ as
\begin{equation}\label{eq:post-sample}
    \bsm_{\text{post}} = \bsm_\text{MAP}^{\bsy,\bsxi} + (I - W_r S_r W_r^T \Gamma_{\text{prior}}^{-1}) \bsm,  
\end{equation}
where $S_r = \text{diag}(s_1, \dots, s_r)$ with $s_j = 1 - 1/\sqrt{\lambda_j + 1}$, $j = 1, \dots, r$, and $\bsm \sim \cN(0,$ $\Gamma_{\text{prior}})$ is a random sample draw from the prior distribution up to a mean term (see more details in \cite{villa2021hippylib}).


\subsection{Computation of the optimality criteria of SBOED}

Efficient computation of the conditional EIG in \eqref{eq:terminalinitialBOED} plays a key role in making the optimization of the sequential experimental design feasible. To this end, we can formulate this computation below based on the Laplace and low-rank approximation of the posterior as presented above. 

First, we draw samples $\bsm_{\text{post}}^{(i-1)}$ from the Laplace approximation of the posterior $\mu(m|$ $\bsy_{1:i-1}^*$, $\bsxi_{1:i-1}^*)$ as in \eqref{eq:post-sample}, with the MAP point $\bsm_{\text{MAP}}^{(i-1)}$ and the posterior covariance $\Gamma_{\text{post}}^{(i-1)}$ computed for the observed data $\bsy_{1:i-1}^*$ at the optimized design $\bsxi_{1:i-1}^*$. We then compute the expectation in \eqref{eq:terminalinitialBOED}, which is given as an integral of the likelihood $\pi(\bsy_{i:K}|m, \bsxi_{1:i:K}, \bsy_{1:i-1}^*)$ with respect to the posterior distribution $\mu(m|\bsy_{1:i-1}^*, \bsxi_{1:i-1}^*)$, i.e.,
\begin{equation}
\pi(\bsy_{i:K}|\bsxi_{1:i:K},\bsy_{1:i-1}^*) = \int_M \pi(\bsy_{i:K}|m, \bsxi_{1:i:K}, \bsy_{1:i-1}^*) d\mu(m|\bsy_{1:i-1}^*, \bsxi_{1:i-1}^*),
\end{equation}

Note that when $i = 1$, we only need to draw samples from the prior distribution. Then we solve the time-dependent PDE \eqref{eq:pde} at these posterior samples, and draw data samples $\bsy_{i:K} = (\bsy_i, \dots, \bsy_K)$ from the noisy observation \eqref{eq:obsdata} corresponding to the experimental design $\bsxi_{i:K}$. 

At each data sample $\bsy_{1:i:K} = (\bsy_1^*, \dots, \bsy_{i-1}^*, \bsy_i, \dots, \bsy_K)$, with the observed data $\bsy_{1:i-1}^*$ from the optimized design $\bsxi_{1:i-1}^*$ and the simulated data $\bsy_{i:K}$ drawn as above from the design $\bsxi_{i:K}$ to be optimized, we can compute the KL divergence in \eqref{eq:terminalinitialBOED} by the Laplace approximation of the posterior $\mu(m| \bsy, \bsxi) \approx \cN(\bsm_{\text{MAP}}^{\bsy, \bsxi}, \Gamma_{\text{post}}^{\bsy, \bsxi})$ in Section \ref{sec:Laplace} with a low-rank approximation of the covariance \eqref{eq:lowrankGamma}, which leads to \cite{wu2023fast, go2023accelerating}
\begin{equation}\label{eq:klapprox_term}
\dkl(\mu(m|\bsy, \bsxi)||\mu(m)) 
\approx 
  \frac{1}{2} \left(
        \sum_{j=1}^r \log (1+\lambda_j) - \frac{\lambda_j}{1+\lambda_j} \right) + \frac{1}{2}||\bsm_{\text{MAP}}^{\bsy, \bsxi} - \bsm_{\text{prior}}||_{ \Gamma_{\text{prior}}^{-1}}^2,
\end{equation}
with the MAP point $\bsm_{\text{MAP}}^{\bsy, \bsxi}$ computed as the solution of the optimization problem \eqref{eq:MAP}, and the eigenvalues $\lambda_j$, $j = 1, \dots, r$, computed as the solution of the generalized eigenvalue problem \eqref{eq:geneig}. Note that we use different data samples $\bsy = \bsy_{1:i:K}$ and $\bsxi = \bsxi_{1:i:K}$ for these computations at different time steps $t_i$. We present the conditional EIG \eqref{eq:terminalinitialBOED} calculation in Algorithm \ref{alg:compute-eig}.

\begin{algorithm}
\caption{Calculation of the conditional EIG in \eqref{eq:terminalinitialBOED} at time $t_i$ for a given $\bsxi_{i:K}$}
\label{alg:compute-eig}
\begin{algorithmic}[1]
\REQUIRE Observed data $\bsy_{1:i-1}^*$ at optimized experimental design $\bsxi_{1:i-1}^*$, the number of data samples $N_s$.  
\ENSURE Conditional EIG in \eqref{eq:terminalinitialBOED}.
\STATE Compute the Laplace approximation of the posterior $\mu(m|\bsy_{1:i-1}^*, \bsxi_{1:i-1}^*) \approx \cN\left(\bsm_{\text{MAP}}^{(i-1)}, \Gamma_{\text{post}}^{(i-1)}\right)$, with $\bsm_{\text{MAP}}^{(i-1)}$ computed by solving \eqref{eq:MAP} and $\Gamma_{\text{post}}^{(i-1)}$ approximated as in \eqref{eq:lowrankGamma} by solving \eqref{eq:geneig}. 
\STATE Initialize the conditional EIG as $\text{cEIG} = 0$, and set $\bsxi = (\bsxi_{1:i-1}^*, \bsxi_{i:K})$.
\FOR{$n = 1$ to $N_s$}
    \STATE Draw a posterior sample of the parameter from $\cN\left(\bsm_{\text{MAP}}^{(i-1)}, \Gamma_{\text{post}}^{(i-1)}\right)$ by sampling from \eqref{eq:post-sample}. 
    \STATE Simulate the system \eqref{eq:pde} at this sample and compute the corresponding data sample $\bsy$ by \eqref{eq:obsdata}.
    \STATE Update $\bsy_{1:i-1}$ with $\bsy_{1:i-1}^*$
    \STATE Solve the optimization problem \eqref{eq:MAP} at $\bsy$ and $\bsxi$ to get the MAP point $\bsm_\text{MAP}^{\bsy,\bsxi}$.
    \STATE Compute the eigenvalues $\lambda_j$, $j = 1,\ldots,r$, by solving \eqref{eq:geneig} at $\bsy$ and $\bsxi$.
    \STATE Compute the information gain (IG) \eqref{eq:klapprox_term} at $\bsy$ and $\bsxi$ and set cEIG = cEIG + IG.
\ENDFOR
\RETURN cEIG = cEIG/$N_s$.
\end{algorithmic}
\end{algorithm}

We remark that the above approximation methods are scalable with respect to the dimension of the parameter space $d_m$ in terms of the number of PDE solves. However, the number of PDE solves may be very large when evaluating and optimizing the optimality criteria of the SBOED, which brings prohibitive computational costs. To address this issue, we propose a deep learning-based surrogate model presented in the next section to approximate the observable $F_k$ and its Jacobian $\nabla_{\bsm} F_k$, $i = 1, \dots, K$, which is further used to approximate the optimality criteria of the SBOED.

\subsection{Adaptive optimization for SBOED}

To solve the adaptive SBOED problem \eqref{eq:terminalinitialBOED}, we follow the process demonstrated in Example \ref{ex:adaptiveSBOED} and present the following adaptive optimization process in Algorithm \ref{alg:adaptive-incre}. 

\begin{algorithm}
\caption{Adaptive optimization for SBOED}
\label{alg:adaptive-incre}
\begin{algorithmic}[1]
\REQUIRE $d$ out of $K$ observation times to be optimized in $d$ steps, and cEIG calculation from Algorithm \ref{alg:compute-eig}.
\ENSURE Optimal observation time $\bsxi^* = (\xi_1^*, \ldots, \xi_K^*)$, where $\xi_i^* \in \{0,1\}$ and $\sum_{k=1}^K \xi_i^* = d$.
\STATE Set $i \gets 1$ (time index after the latest observation)
\STATE Initialize $\bsxi_{1:i:K} \in \bR^{K}$ and $\bsy_{1:i:K} \in \bR^{d_y \times K}$, e.g., both as zeros.
\FOR{step $= 1$ to $d$}
    \STATE Solve the optimization problem \eqref{eq:terminalinitialBOED} for the optimal experimental design $\bsxi_{i:K}^*$. \label{step:solve}
    \STATE Set $i = \arg\min_{j \in  i:K} \xi_j^* = 1$ in $\bsxi_{i:K}^*$, the first time index with nonzero design. 
    \STATE Progress the dynamical system until time $t_i$ and make observation of real data $\bsy_i^*$.
    \STATE Set $i = i+1$ and update $\bsxi_{1:i:K} \in \bR^{K}$ and $\bsy_{1:i:K} \in \bR^{d_y \times K}$ with  $\bsxi_{1:i-1}^*$ and $\bsy_{1:i-1}^*$.
\ENDFOR
\RETURN $\bsxi^*$
\end{algorithmic}
\end{algorithm}

We remark that to solve the optimization problem \eqref{eq:terminalinitialBOED} in line \ref{step:solve} of Algorithm \ref{alg:adaptive-incre}, we can loop through all the possible combinations of experimental design for the remaining observation times, compute the conditional EIG corresponding to each combination, and select the one with the largest conditional EIG. This brute force combinatorial optimization is feasible when $d$ and/or $K$ are small. When they become very large, we can apply a greedy algorithm to select $\bsxi_{i:K}^*$ as in \cite{foster2021deep, ivanova2021implicit} in each step or multiple steps forward \cite{murphy2003optimal} to choose to reduce computational cost at the expense of potentially not finding the globally optimal solution. 

\section{Derivative-informed latent attention neural operator}\label{sec:LANO}

In this section, we introduce a novel neural network surrogate model to approximate both the PtO maps and their Jacobians at given time steps of the dynamical system, which are used to compute the optimality criteria of the SBOED. This surrogate model integrates dimension reduction of the parameter and observable to the latent space, an attention-based architecture to capture the temporal correlation of the latent dynamics, and derivative-informed training of the neural network, which together achieve high accuracy, efficiency, and scalability of the approximation for both the PtO maps and their Jacobians, and for the optimality criteria.

\subsection{Derivative-informed dimension reduction}
\label{sec:dim-red}
 
As the dimensions of the input parameters and the output observables are very high in our case, we first employ dimension reduction to compress the input and output to low-dimensional subspaces to construct a parsimonious neural network approximation of the nonlinear mapping between the low-dimensional subspaces. For computational efficiency and convenience in evaluating both the PtO map and its Jacobian, we use linear dimension reduction methods, including the Jacobian/derivative-informed input subspace (DIS) and principal component analysis (PCA) for output dimension. 




For the input dimension reduction, we use linear projection with bases of  derivative-informed input subspace (DIS) or active subspace, which has been shown as one of the most effective linear reduction methods in Bayesian inverse problems \cite{zahm2022certified, chen2020projected, wang2022projected} and Bayesian optimal experimental design problems \cite{go2023accelerating, wu2023large}.  
In the setting of the dynamical system and observations, we compute the bases as the eigenvectors of the following generalized eigenvalue problem with the cumulative Jacobian information
\begin{equation}\label{eq:eigenias}
    \E_{\bsm} \left[\sum_{k=1}^K\nabla_{\bsm} F_k^T(\bsm)\nabla_{\bsm} F_k(\bsm)\right]\psi_{\bsm}^{(i)} = \lambda_i\Gamma_\text{prior}^{-1}\psi_{\bsm}^{(i)}, \quad i = 1, ..., r_m,
\end{equation}
where $\psi_{\bsm}^{(i)}$ are the generalized eigenvectors, $\lambda_i$ are the $r_m$ largest generalized eigenvalues with $\lambda_1 \geq \cdots \geq \lambda_{r_m}$ and $({\psi}_{\bsm}^{(i)})^T{\Gamma}_\text{prior}^{-1}({\psi}_{\bsm}^{(j)})=\delta_{ij}$. The generalized eigenvalue problem \eqref{eq:eigenias} can be solved by randomized algorithm \cite{villa2021hippylib}, where the expectation can be evaluated by sample average approximation with $N_\bsm$ samples, and the action of $\nabla_\bsm F_k^T \nabla_\bsm F_k$ in a given direction can be computed as in Appendix \ref{ap:Jac}.
Let $\Psi_{\bsm} = (\psi_{\bsm}^{(1)}, \ldots, \psi_{\bsm}^{(r_m)})$ denote the projection bases,
the input parameter $\bsm$ can then be approximated as
\begin{equation} \label{eq:DIS}
\bsm \approx \bsm_r := \bsm_\text{prior} + \Psi_{\bsm} \dbm,
\end{equation}
where $\dbm = \Psi_{\bsm}^T \Gamma_{\text{prior}}^{-1} (\bsm -  \bsm_{\text{prior}}) \in \bR^{r_\bsm}$ is the projection coefficient vector. 


For the output dimension reduction, we use a common PCA. We first concatenate the observables across all $K$ time steps and $N_t$ samples into a snapshot matrix $\mathbb{B} = [F_1^{(1)}, \ldots,F_K^{(1)},...,F_1^{(N_t)}, \ldots,F_K^{(N_t)}]$. We then compute the sample mean $\bar{F}$ and perform a truncated SVD on the centered data matrix $\hat{\mathbb{B}} = \mathbb{B}-\bar{F}$ as  
\begin{equation} \label{eq:PCA}
\hat{\mathbb{B}} \approx \hat{\mathbb{B}}_r := {\Psi}_{F} {\Sigma}_{F} {\Phi}_{F}^T.
\end{equation}
Here, ${\Psi}_{F} = [\psi_{F}^{(1)}, \dots, \psi_{F}^{(r_F)}]$ and ${\Phi}_F = [\phi_{F}^{(1)}, \dots, \phi_{F}^{(r_F)}]$ contain the first $r_F$ left and right singular vectors corresponding to the $r_F$ largest singular values $\sigma_1 \geq \dots \geq \sigma_{r_F}$ with $\Sigma_{F} = \text{diag}(\sigma_1, \dots, \sigma_{r_F})$.

Then the observable $F_k$ at time $t_k$ can be approximated by linear projection to the bases $\Psi_F$ as 
\begin{equation}
F_k \approx F_{k}^r := \bar{F} + {\Psi}_{F} \beta_{F_k}, \quad k = 1, \dots, K,
\end{equation}
where $\beta_{F_k} = {\Psi}_{F}^T(F_k - \bar{F}) \in \bR^{r_F}$ is the projection coefficient vector. 


\subsection{Latent attention neural operator}
\label{sec:nn}

In addition to the linear dimension reduction, efficient use of training data is crucial due to its high cost. Studies in \cite{hestness2017deep} and \cite{kaplan2020scaling} have shown that larger models can perform better with increasing training samples, but may overfit with insufficient data. Larger neural networks offer strong expressibility and high accuracy, but require substantial training data. Conversely, smaller networks need fewer training samples but may lack accuracy for state prediction. To address this trade-off, we propose a neural network architecture that minimizes required training data while maintaining accuracy for SBOED applications.

Our approach draws inspiration from successful sequential models, particularly attention models \cite{vaswani2017attention} for their strong performance in sequential tasks, and latent dynamics models \cite{regazzoni2023latent} for their ability to efficiently train dynamics in low-dimensional latent variables. Based on these insights, our proposed neural network comprises two main components: 1) an attention layer to capture temporal dependence, and 2) latent dynamics to train dynamics in the reduced latent variables. This architecture aims to balance computational efficiency with the ability to capture complex sequential relationships in PDEs. Furthermore, we design the network to simultaneously predict the evolution of states and their corresponding Jacobians.

To this end, we propose the following neural network architecture with four components: 1) latent encoding to encode the input and output to a latent space, 2) latent attention to use the attention mechanism to learn the latent dependence, 3) latent dynamics to model the dynamics in the latent variables with attention, and 4) latent decoding to decode the latent dynamics to the observable dynamics. We call this neural network a latent attention neural operator (LANO). 

\begin{enumerate}
    \item \textbf{Latent encoding}: Given input and output data pairs $(\bsm, \bsF)$, with the parameter $\bsm$ and the observables $\bsF = (F_0(\bsm), F_1(\bsm), \dots F_K(\bsm))$, we first use the linear dimension reduction DIS and PCA in Section \ref{sec:dim-red} to compute the reduced representation $\beta_\bsm \in \bR^{r_\bsm}$ and $\beta_\bsF = (\beta_{F_0}, \beta_{F_1}, \dots, \beta_{F_K}) \in \bR^{r_F \times (K+1)}$, and then apply a linear transformation layer for both of them as 
    \begin{equation}
        p = W^p \beta_\bsm + b^p \; \text{ and } \; 
        s_k = W_k^s \beta_{F_{k}} + b_k^s, \quad k = 0, \dots, K-1,
    \end{equation}
    with the learnable neural network parameters $W^p \in \bR^{d_h \times r_\bsm}$, $b^p \in \bR^{d_h}$, $W_k^s \in \bR^{d_h \times r_F}$, and $b_k^s \in \bR^{d_h}$, for a hidden latent dimension $d_h$. The output of the transformed state $s_k$ and transformed parameter $p$ are then concatenated as $(s_k; p) \in \bR^{2d_h}$, which is encoded to a latent variable $z_k$ at time $t_k$ as 
    \begin{equation}
        z_k = \sigma_z (W_k^z (s_k;p) + b_k^z), \quad k = 0, \dots, K-1,
    \end{equation}
    with learnable $W_k^z \in \bR^{d_h \times 2d_h}$ and $b_k^z \in \bR^{d_h}$, and an activation function $\sigma_z$, e.g., tanh. This latent encoding is motivated by the fact that the PDE system \eqref{eq:pde} depends on the state and the parameter at each time step, both of which allow low-dimensional representation by compression.
    
    \item \textbf{Latent attention}: We denote $Z = (z_0, \dots, z_{K-1}) \in \bR^{d_h \times K}$ as the aggregated latent variable. We apply an attention layer by first computing the query, key, and value matrices as 
    \begin{equation}
        \cQ = Z^T W_\cQ, \quad \cK = Z^T W_\cK, \quad \cV = Z^T W_\cV,
    \end{equation}
    with learnable $W_\cQ, W_\cK, W_\cV \in \bR^{d_h \times d_a}$, and an attention dimension $d_a$. In practice, we can add a positional encoding to $Z$ before computing these quantities, allowing the naturally permutation-invariant attention layer to respect positioning in $Z$ as in \cite{vaswani2017attention}. We then compute the attention as
    \begin{equation}
        \cA = \text{softmax}\left(\frac{\cQ\cK^T}{\sqrt{d_a}} + \cM_F\right)\cV
    \end{equation}
    where $\cM_F \in \mathbb{R}^{K \times K}$ represents a lower triangular mask for the attention $\cA \in \bR^{K\times d_a}$, ensuring causal dependency of the latent variable in time. 
    
    \item \textbf{Latent dynamics}: To model the latent dynamics of the latent variable with attention, we first transform the attention to the latent space by two layer neural networks 
    \begin{equation}\label{eq:fFfJ}
    \begin{split}
        f & = \sigma_f (\cA W_1 + b_1) W_2 + b_2 \in \bR^{K \times d_h},
    \end{split}
    \end{equation}
    with learnable $W_1 \in \bR^{d_a \times d_h}$, $W_2 \in \bR^{d_h \times d_h}$, and $b_1, b_2 \in \bR^{d_h}$, and an activation function $\sigma_f$, e.g., ELU in \cite{clevert2015fast} to allow sufficient derivative information. Then, we apply a layer normalization to maintain stable activation. Finally, we construct two implicitly dependent latent dynamics to learn the latent variable $\beta_\bsF = (\beta_{F_0}, \beta_{F_1}, \dots, \beta_{F_K})$, with $\beta_{F_0}$ given as the initial condition, using ResNet layers \cite{he2016deep} as 
    \begin{equation}
    \begin{split}
        \beta^F_{{k+1}} &= \beta^F_{{k}} + W_{2,k}^F \sigma_\beta (W_{1,k}^F f_k + b_{1,k}^F) + b_{2,k}^F, \\
        \beta^J_{{k+1}} & = \beta^J_{{k}} + W_{2,k}^J \sigma_\beta (W_{1,k}^J f_k + b_{1,k}^J) + b_{2,k}^J,
    \end{split}
    \end{equation}
    where $f_k$ is (the transpose of) the $k$-th rows of $f$ in \eqref{eq:fFfJ}, $\sigma_\beta$ is an activation function, e.g., ELU, and $W_{1,k}^F, W_{1,k}^J \in \bR^{r_F \times d_h}$, $W_{2,k}^F, W_{2, k}^J \in \bR^{r_F \times r_F}$, and $b_{1, k}^F, b_{2, k}^F, b_{1, k}^J, b_{2, k}^J \in \bR^{r_F}$ are learnable parameters for each time step $k = 0, \dots, K-1$. We set the initial condition $\beta_0 = \beta_{F_0}$. 
    \item \textbf{Latent decoding}: At the final step, we decode the latent variables to the full space by PCA as 
    \begin{equation}
        \begin{split}
            \hat{F}_k & = \bar{F} + \Psi_F \beta^F_{k},\\
            \hat{J}_k & = \nabla_\bsm (\bar{F}+ \Psi_F \beta^J_k),
        \end{split}
        \label{eq:loss-train}
    \end{equation}
    for $k = 1, \dots, K$, where $\hat{F}_k$ and $\hat{J}_k$ are the neural network approximations of the observation $F_k$ and its Jacobian $J_k = \nabla_\bsm F_k$, with the derivative $\nabla_\bsm $ in $\hat{J}_k$ computed using automatic differentiation.
\end{enumerate}
In a compact form, we denote the neural network approximations of the observable $\bsF=(F_1, \dots, F_K)$ and its Jacobian $\bsJ = \nabla_\bsm \bsF$ in the Encoder--Neural Network--Decoder format  
\begin{equation}\label{eq:ENND}
    \begin{split}
        \hat{F}_\bstheta(\bsm) & = \cD_{\Psi_F} \circ \cN_{\bstheta}^F \circ \cE_{\Psi_\bsm} (\bsm),\\
        \hat{\bsJ}_\bstheta(\bsm) & = \Psi_F \, \nabla_{\beta} \cN_{\bstheta}^J(\beta_\bsm) \, \Psi_\bsm^T \, \Gamma_{\text{prior}}^{-1},
    \end{split}
\end{equation}
where $\cN_\bstheta^F: \bR^{r_\bsm} \to \bR^{r_F \times K}$ and $\cN_\bstheta^J: \bR^{r_\bsm} \to \bR^{r_F \times K}$ represent the neural network approximations with learnable parameters $\bstheta$,  $\cE_{\Psi_\bsm} : \bR^{d_\bsm} \to \bR^{r_\bsm}$ is an encoder defined by the linear projection \eqref{eq:DIS} with basis $\Psi_\bsm \in \bR^{d_\bsm \times r_\bsm}$ as $\cE_{\Psi_\bsm}(\bsm) = \beta_\bsm = \Psi_{\bsm}^T \Gamma_{\text{prior}}^{-1} (\bsm -  \bsm_{\text{prior}})$, 
and $\cD_{\Psi_F}: \bR^{r_F} \to \bR^{d_F}$ is a decoder defined by the linear projection \eqref{eq:PCA} with basis $\Psi_F \in \bR^{d_F \times r_F}$, with $\cD_{\Psi_F}(\beta) = \bar{F} + \Psi_F \beta$ for any $\beta \in \bR^{r_F}$.

Key features of this architecture include 1)~a causal attention mechanism, which allows the network to capture causal relationships for forward prediction and Jacobian computation; 2)~latent dynamics layers, which process the reduced-dimension representations, enabling the network to learn complex, nonlinear relationships in the reduced space; and 3)~automatic differentiation, which is used to compute the Jacobians efficiently, reducing computational cost compared to traditional methods.

Combining these elements enables the network to handle the complexities of a PDE-based model in the reduced space. This approach offers several advantages: 1)~computational efficiency, as it works in a reduced-order space and uses automatic differentiation to handle complex systems more efficiently than full-order models; 2)~simultaneous learning, where the network learns to predict state evolution and compute Jacobians in a single framework, potentially capturing intricate relationships between the two tasks; and 3)~flexibility, as the architecture can be adapted to various PDEs by adjusting the dimensionality reduction techniques.

\subsection{Data generation and derivative-informed training}
\label{sec:data}

To train the neural operator of latent attention, we use both the observable $\bsF(\bsm) = (F_1(\bsm), \dots, F_K(\bsm))$ and its Jacobian $\bsJ(\bsm) = \nabla_\bsm \bsF(\bsm)$ as targets to match by the neural network approximations for the training data. To generate the training data, we first draw $N_t$ samples of $\bsm^{(n)}$, $n = 1, \dots, N_t$, from its prior distribution. For each sample $\bsm^{(n)}$, we solve the PDE \eqref{eq:pde} to obtain the full-space observations $\bsF(\bsm^{(n)}) = (F_1(\bsm^{(n)}), \dots, F_K(\bsm^{(n)}))$, for $k=1,\ldots, K$. We then apply dimension reduction techniques (DIS and PCA, as detailed in Section \ref{sec:dim-red}) to project the high-dimensional input parameters and observations into the reduced spaces, i.e., $\bsm^{(n)} \rightarrow \beta_\bsm^{(n)}$ and $\bsF(\bsm^{(n)}) \rightarrow \beta_\bsF^{(n)}$. Additionally, we compute the reduced Jacobian $\beta_\bsJ^{(n)} = \Psi_{F}^T \,  \bsJ^{(n)} \, \Psi_{\bsm}$, a projection of the full Jacobian in both input and output spaces, which only requires solving $\min(r_\bsm,r_F)$ linearized PDEs, as presented in Appendix \ref{ap:Jac}.

With the data set $(\beta_\bsm^{(n)}, \beta_\bsF^{(n)}, \beta_\bsJ^{(n)})$, $n = 1, \dots, N_t$, all computed in the reduced dimensions, we define the derivative-informed empirical loss function to train LANO as 
\begin{equation}\label{eq:loss}
    \ell(\bstheta) = \sum_{n=1}^{N_t} 
    ||\beta_\bsF^{(n)} - \cN_\bstheta^F(\beta_\bsm^{(n)})||^2 + ||\beta_\bsJ^{(n)} - \nabla_{\beta} \cN_\bstheta^J(\beta_\bsm^{(n)})||^2,
\end{equation}
whose evaluation and optimization are made efficient as all the quantities are relatively small depending only on the reduced dimensions $r_\bsm$ and $r_F$, not the full dimensions $d_\bsm \gg r_\bsm$ and $d_F \gg r_F$.
This loss function balances the accuracy of state evolution prediction with the accuracy of Jacobian computation, enabling the network to learn both tasks simultaneously. Note that this derivative-informed training is inspired by DINO \cite{o2024derivative} and differs in that the same neural network is trained in DINO, while two neural networks for the output and its Jacobian are trained separately in LANO to achieve a balanced accuracy of the two terms.

\subsection{Efficient computation of the optimality criteria for SBOED}

In the computation of the conditional EIG by Algorithm \ref{alg:compute-eig}, which is the optimality criteria of the adaptive SBOED problem \eqref{eq:terminalinitialBOED}, we need to  
\begin{enumerate}
    \item compute the MAP points $\bsm_{\text{MAP}}^{(i-1)}$ and $\bsm_{\text{MAP}}^{\bsy, \bsxi}$ by solving the optimization problem \eqref{eq:MAP},
    \item compute the eigenvalues of the generalization eigenvalue problem \eqref{eq:geneig},
    \item draw samples from the posterior $\cN(\bsm_{\text{MAP}}^{(i-1)}, \Gamma_{\text{post}}^{(i-1)})$ by \eqref{eq:post-sample}, 
    \item simulate the dynamical system \eqref{eq:pde} at these samples. 
\end{enumerate}
All these steps are very expensive and involve solving high-fidelity optimization problems, generalized eigenvalue problems, sampling, and simulation many times. In this section, we present efficient computation using neural network approximations to accelerate all these steps significantly. The simulation in step 4 can be directly replaced by the neural network approximation in \eqref{eq:ENND}. We present the first three steps below.

\subsubsection{Computing the MAP point.}\label{eq:reduced-MAP}  
Once trained with the loss function \eqref{eq:loss}, the neural network approximations can be used to compute the MAP point in \eqref{eq:MAP} as $\bsm_{\text{MAP}}^{\bsy, \bsxi} = \bsm(\beta_{\text{MAP}}^{\bsy, \bsxi})$ using \eqref{eq:DIS} for the reduced MAP point $\beta_{\text{MAP}}^{\bsy, \bsxi}$ by solving
\begin{equation} \label{eq:inv-red}
    \beta_{\text{MAP}}^{\bsy,\bsxi} = \arg\min_{\beta \in \bR^{r_\bsm}} \frac{1}{2}\sum_{k=1}^K \xi_k\|\bsy_k - \Psi_F \, (\cN_\bstheta^F(\beta))_k\|_{\Gamma_{\text{noise}}^{-1}}^2 + \frac{1}{2}\|\beta\|^2_{\Gamma_{\dbm}^{-1}},
\end{equation}
where $(\cN_\bstheta^F(\beta))_k \in \bR^{r_F}$ is the $k$-th output of the neural network at time $t_k$, $\Gamma_{\dbm} = \Psi_{\bsm}^T\Gamma_\text{prior}^{-1}\Psi_{\bsm} = I$, which is identity, as the DIS bases $\Psi_\bsm$ are orthonormal with respect to $\Gamma_{\text{prior}}^{-1}$. This optimization problem is in the reduced space of small dimension $r_\bsm$, which can be efficiently solved by a gradient-based method using automatic differentiation with respect to $\beta$.  

\subsubsection{Solving the generalized eigenvalue problem.}\label{eq:reduced-eigen} 
In the computation of the eigenpairs of the generalized eigenvalue problem \eqref{eq:geneig} with the Gauss--Newton approximation of the Hessian given in \eqref{eq:GN}, we need to evaluate the Jacobian at the MAP point $\nabla_{\bsm}F_k(\bsm_\text{MAP}^{\bsy,\bsxi})$ for $k = 1, \dots, K$. Note that this can be evaluated by the neural network approximation $\hat{\bsJ}_\bstheta$ in \eqref{eq:ENND}. 
We use this approximate Jacobian in the generalized eigenvalue problem \eqref{eq:geneig}, approximating the eigenvectors by $\bsw_j = \Psi_\bsm \bsu_j$ and left multiplying $\Psi_\bsm^T$ on both sides of \eqref{eq:geneig}, which leads to the reduced eigenvalue problem
\begin{equation}\label{eq:red-geigen}
\hat{H}_{\text{misfit}}^{\bsy,\bsxi}(\beta_\text{MAP}^{\bsy,\bsxi}) \bsu_j = \lambda_j \bsu_j, \quad j = 1, \ldots, r_\bsm,
\end{equation}
where the reduced matrix $\hat{H}_{\text{misfit}}^{\bsy,\bsxi}(\beta_\text{MAP}^{\bsy,\bsxi}) \in \mathbb{R}^{r_\bsm \times r_\bsm}$ is given by
\begin{equation}
\hat{H}_{\text{misfit}}^{\bsy,\bsxi}(\beta_\text{MAP}^{\bsy,\bsxi}) = \sum_{k=1}^K\xi_k(\nabla_\beta\cN_\bstheta^J(\beta_{\text{MAP}}^{\bsy,\bsxi}))_k^T \Psi_{F}^T \Gamma_{\text{noise}}^{-1} \Psi_{F} (\nabla_\beta\cN_\bstheta^J(\beta_{\text{MAP}}^{\bsy,\bsxi}))_k,
\end{equation}
which can be efficiently computed with the reduced Jacobian at time step $k$ as $(\nabla_\beta\cN_\bstheta^J(\beta_{\text{MAP}}^{\bsy,\bsxi}))_k \in \bR^{r_F \times r_\bsm}$.   

Note that with the MAP point and the eigenvalues, we can evaluate the information gain in \eqref{eq:klapprox_term} as 
\begin{equation}\label{eq:reduced-klapprox_term}
\dkl(\mu(m|\bsy, \bsxi)||\mu(m)) 
\approx 
  \frac{1}{2} \left(
        \sum_{j=1}^{r_\bsm} \log (1+\lambda_j) - \frac{\lambda_j}{1+\lambda_j} \right) + \frac{1}{2}||\beta_{\text{MAP}}^{\bsy, \bsxi}||^2.
\end{equation}

\subsubsection{Sampling from the Laplace approximation.}
Given observation data $\bsy_{1:i-1}^*$ at optimized experimental design $\bsxi_{1:i-1}^*$, to draw the posterior samples from the Laplace approximation 
$\cN(\bsm_{\text{MAP}}^{(i-1)}, \Gamma_{\text{post}}^{(i-1)})$ by \eqref{eq:post-sample}, we first solve for the MAP point $\beta_{\text{MAP}}^{(i-1)}$ as in Section \ref{eq:reduced-MAP} and compute the eigenpairs $(\lambda^{(i-1)}_j, \bsu^{(i-1)}_j)$, $j = 1, \dots, r $ with $r = r_\bsm$ as in Section \ref{eq:reduced-eigen}. Then we draw the posterior samples of $\beta_\text{post}^{(i-1)}$ as the input of the neural networks $\cN_\bstheta^F$ and $\cN_\bstheta^J$ for the simulation of the system. This is given as the projected coefficient vector of the posterior sample in \eqref{eq:post-sample} as 
\begin{equation}\label{eq:reduced-post-sample}
    \beta^{(i-1)}_{\text{post}} = \beta_{\text{MAP}}^{(i-1)} + (I_r - U^{(i-1)}_r S_r^{(i-1)} (U_r^{(i-1)})^T) \beta,
\end{equation}
where $U_r^{(i-1)} = (\bsu_1^{(i-1)}, \dots, \bsu_r^{(i-1)}) \in \bR^{r_\bsm \times r_\bsm}$, 
and $\beta \sim \cN(0, I_r)$ with identity $I_r \in \bR^{r_\bsm \times r_\bsm}$. We establish \eqref{eq:reduced-post-sample} from \eqref{eq:post-sample} by replacing in the right hand side of \eqref{eq:post-sample} the following quantities: the MAP point $\bsm_{\text{MAP}}^{(i-1)} \approx \bsm_{\text{prior}} + \Psi_\bsm \beta_{\text{MAP}}^{(i-1)}$, the eigenvectors $W_r^{(i-1)} \approx \Psi_\bsm U_r^{(i-1)}$, and the random sample drawn from the prior distribution up to a mean term $\bsm = \Gamma_{\text{prior}}^{1/2} \eta \approx \Psi_\bsm \beta$ with $\eta \sim \cN(0, I)$ for identity $I \in \bR^{d_\bsm \times d_\bsm}$ and $\beta = \Psi_\bsm^T \Gamma_{\text{prior}}^{-1}\bsm$ by projection, which leads to 
\begin{equation}
\begin{split}
\bsm_{\text{post}}^{(i-1)} & = \bsm_\text{MAP}^{(i-1)} + (I - W_r^{(i-1)} S_r^{(i-1)} (W_r^{(i-1)})^T \Gamma_{\text{prior}}^{-1})\bsm \\
& \approx \bsm_{\text{prior}} + \Psi_\bsm \beta_{\text{MAP}}^{(i-1)} + (I - \Psi_\bsm U_r^{(i-1)} S_r^{(i-1)} (U_r^{(i-1)})^T \Psi_\bsm^T \Gamma_{\text{prior}}^{-1}) \Psi_\bsm \zeta \\
& = \bsm_{\text{prior}} + \Psi_\bsm \beta^{(i-1)}_{\text{post}},
\end{split}
\end{equation}
which implies that $\beta^{(i-1)}_{\text{post}}$ is the projected coefficient vector of $\bsm_{\text{post}}^{(i-1)}$ by the DIS projection. Finally, we note that the covariance of $\zeta$ is given by 
\begin{equation}
    \bE[\beta \beta^T] = \Psi_\bsm^T \Gamma_{\text{prior}}^{-1} \bE[\bsm \bsm^T] \Gamma_{\text{prior}}^{-1} \Psi_\bsm = \Psi_\bsm^T  \Gamma_{\text{prior}}^{-1} \Psi_\bsm = I_r,
\end{equation}
where we have $\bE[\bsm \bsm^T] = \Gamma_{\text{prior}}$, so that $\beta \sim \cN(0, I_r)$.

\subsection{Computational complexity}
\label{sec:complexity}

In this subsection, we analyze and compare the computational cost of FEM and the proposed surrogate LANO in solving the SBOED problem. We use the same optimization Algorithm \ref{alg:adaptive-incre} to solve the SBOED problem of adaptively selecting $d$ observation times from $K$ candidate times using the adaptive terminal formulation \eqref{eq:terminalinitialBOED}. The optimality criteria of the conditional EIG in Algorithm \ref{alg:adaptive-incre} is computed by Algorithm \ref{alg:compute-eig} for $N_{\text{opt}}$ times, which is upper bounded by 
$N_{\text{opt}} \leq N_{\text{max}} = {K \choose d} + {K-1 \choose d-1} +\dots + {K-d+1 \choose 1} = {K+1 \choose d}$.
Each conditional EIG evaluation requires computing the information gain $N_s$ times \eqref{eq:klapprox_term} by sample average approximation with $N_s$ samples, which leads to a total of $N_{\text{opt}} N_s$ times evaluation of the information gain.

The acceleration of the LANO surrogate compared to the FEM  comes from the computation of (1) the MAP point in \eqref{eq:MAP} by FEM vs in \eqref{eq:inv-red} by LANO, (2) the eigenpairs in \eqref{eq:geneig} by FEM and in \eqref{eq:red-geigen} by LANO,  (3) the sampling from the Laplace approximation of the posterior in \eqref{eq:post-sample} by FEM and in \eqref{eq:reduced-post-sample} by LANO, and (4) the information gain in \eqref{eq:klapprox_term} by FEM and in \eqref{eq:reduced-klapprox_term} by LANO. Once the MAP point and the eigenpairs are computed, the cost for sampling from the posterior and the evaluation of the information gain are negligible for both FEM and LANO. Therefore, we focus on the analysis of (1) and (2) in terms of the number of PDE solves by FEM, which dominate the total computation for large-scale PDE models. For comparison, we analyze the cost in the number of PDE solves for the offline construction of the LANO surrogate.

Let $C_1$ denote the cost in solving the (possibly nonlinear) state PDE \eqref{eq:pde} (e.g., by a discretization in the form of \eqref{eq:state-t}), and let $C_2$ denote the cost in solving the linearized PDE \eqref{eq:hatu} or \eqref{eq:phat} in the computation of the directional derivatives. As the linear operators in the linearized PDE is the same for the derivative acting in different directions, we can amortize the solve by factorizing the linear operators (e.g., by LU factorization) and use the factorizations to solve the linearized PDE many times, which may lead to $C_2 \ll C_1$.

The cost for each evaluation of the information gain by FEM in \eqref{eq:klapprox_term} is dominated by one solve of the optimization problem \eqref{eq:MAP} to compute the MAP point and one solve of the generalized eigenvalue problem \eqref{eq:geneig} to compute the eigenpairs. By an inexact Newton-CG
algorithm \cite{villa2021hippylib}, with $N_{nt}$ Newton iterations and $N_{cg}$ CG iterations (in average) per Newton iteration, we can solve $N_{nt}$ times the state PDE \eqref{eq:pde} with a cost of $N_{nt} C_1$ and $2N_{nt} N_{cg}$ times the linearized PDEs (each Hessian action require 2 linearized PDE solves) with a cost of $2N_{nt} N_{cg} C_2$. By a double pass randomized algorithm, we can solve the generalized eigenvalue problem \eqref{eq:geneig} by one state PDE solve at the MAP point with a cost of $C_1$, and $4(r_e+p)$ linearized PDE solves with a cost of $4(r_e + p) C_2$, where $r_e$ is the number of eigenpairs and $p$ is an oversampling parameter \cite{villa2021hippylib}, e.g., $p = 5$. We report the dominate cost for solving the SBOED problem in computing the MAP point and eigenpairs for $N_{\text{opt}}N_s$ times by FEM in Table \ref{tb:comp}.



For the offline training of LANO, we need to compute the input and output dimension reduction bases and generate training data, for which the cost of PDE solves are dominate. Specifically, we first solve $N_t$ state PDEs to compute the PtO map at $N_t$ training samples with a cost of $N_t C_1$. Then we compute $r_\bsm$ input DIS bases using $N_\bsm < N_t$ training samples, with an additional cost of $4 N_\bsm (r_\bsm+p) C_2$ to solve $4 N_\bsm (r_\bsm+p)$ linearized PDEs for Jacobian actions in \eqref{eq:eigenias}. We compute the $r_F$ output PCA bases using $N_F < N_t$ training samples by truncated SVD without solving additional PDEs. Finally, for each training sample, we compute the reduced Jacobian in Section \ref{sec:data} with an additional cost of $r_t C_2$ with $ r_t = \min(r_\bsm, r_F)$ in solving $r_t$ linearized PDEs. See Table \ref{tb:comp} for a summary of the offline data generation cost, and Table \ref{tab:ct_nn_fem_2} for the offline training cost and Table \ref{tab:comparison} for the online evaluation cost of LANO compared to FEM for a specific example.

\begin{table}[h]
  \footnotesize
  \begin{center}
    \begin{tabular}{|c|c||c|c|}
      \hline
       cost & FEM & offline cost & LANO \\ 
      \hline
      MAP point &  $N_{\text{opt}} N_s N_{nt} (C_1 + 2N_{cg} C_2)$ & 
      Training data & $N_t(C_1 + r_tC_2)$ \\
      \hline
      Eigenpairs & $N_{\text{opt}} N_s(C_1 + 4(r_e+p)C_2)$ & DIS bases & $ 4 N_\bsm (r_\bsm+p)C_2$ \\ 
      \hline
    \end{tabular}
    \caption{Computational complexity in terms of the cost for PDE solves, with a cost $C_1$ to solve one state PDE and $C_2$ to solve one linearized PDE. $N_{\text{opt}}$: \# evaluations of the conditional EIG, $N_s$: \# samples to compute each conditional EIG, $N_{nt}$: \# Newton iterations, $N_{cg}$: \# CG iterations per Newton iteration, $r_e$: \# eigenpairs, $p$: \# oversampling parameter, $N_t$: \# training samples, $N_\bsm < N_t$: \# samples to compute input DIS bases, $r_t = \min(r_\bsm, r_F)$ with $r_\bsm$ input DIS bases and $r_F$ output PCA bases.}\label{tb:comp}
  \end{center}
  \end{table}

\section{Numerical experiment}\label{sec:numerics}
In this section, we conduct experiments to demonstrate the performance of our proposed computational framework, applying it to sequential optimal design of the time to take images using MRI to infer tumor growth. 
Specifically, we focus on glioblastoma, the most aggressive primary brain tumor. Medical imaging techniques often struggle to identify the boundary of the tumor precisely, potentially leading to suboptimal interventions and prognoses \cite{stupp2005radiotherapy, liang2023image, liang2023bayesian}. In clinical practice, obtaining daily MRI images (e.g., over ten days) would provide the most comprehensive information for treatment planning. However, this approach is time-consuming and expensive. Identifying the most informative time points for MRI imaging can be approached as a sequential experimental design problem. 

\subsection{Setup of the tumor growth model}
To evaluate the performance of our proposed method, we utilize the brain tumor model presented in \cite{liang2023bayesian} to select the optimal imaging time in the context of SBOED. The model of the proliferation and infiltration of the tumor growth is described by a reaction-diffusion equation with a nonlinear reaction term 
\begin{equation}
\begin{split}
    \frac{\partial u}{\partial t} = \nabla \cdot (D \nabla u) + G (1 - u) u& \ \ \ \ \text{ in } \Omega \times (0, T],\\
    D \nabla u \cdot n = 0& \ \ \ \ \text{ on } \partial \Omega \times (0, T],\\
    u(x,0) = u_0& \ \ \ \ \text{ in } \Omega,
\end{split}
\end{equation}
where $\Omega$ denotes the brain domain of a specific rat extracted from a segmented 2D slice of a $T_2$-weighted MRI image and the function $u(x,t) \in [0,1]$ quantifies the estimated tumor volume fraction at position $x$ and time $t$. We use a homogeneous Neumann boundary condition and set the initial condition as $u_0$. 

\begin{figure}[!htb]
    \centering
    \includegraphics[width=0.26\textwidth]{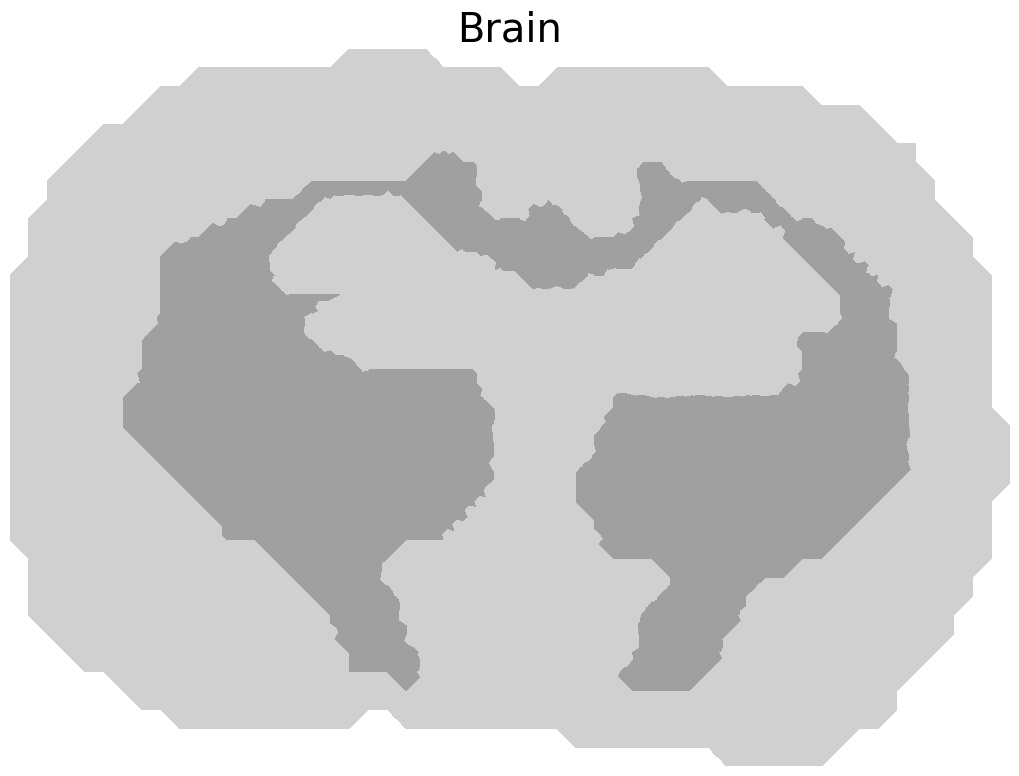}
    \hspace{0.02\textwidth}
    \includegraphics[width=0.31\textwidth]{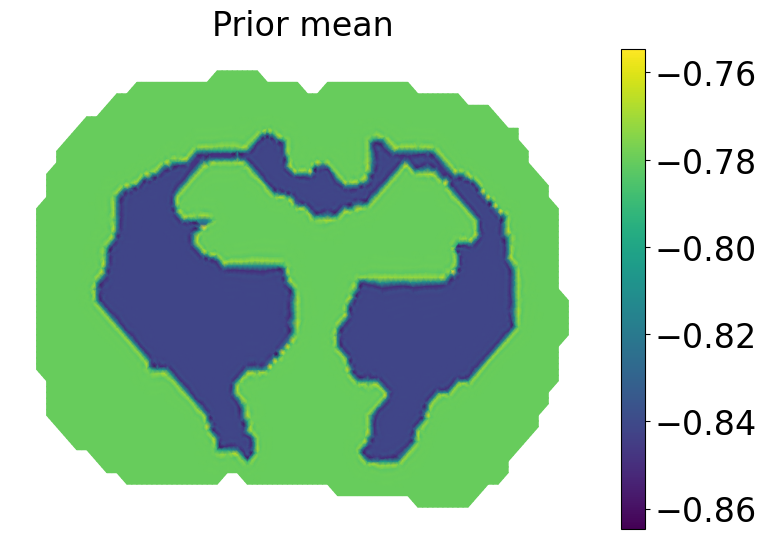}
    \hspace{0.02\textwidth}    
    \includegraphics[width=0.30\textwidth]{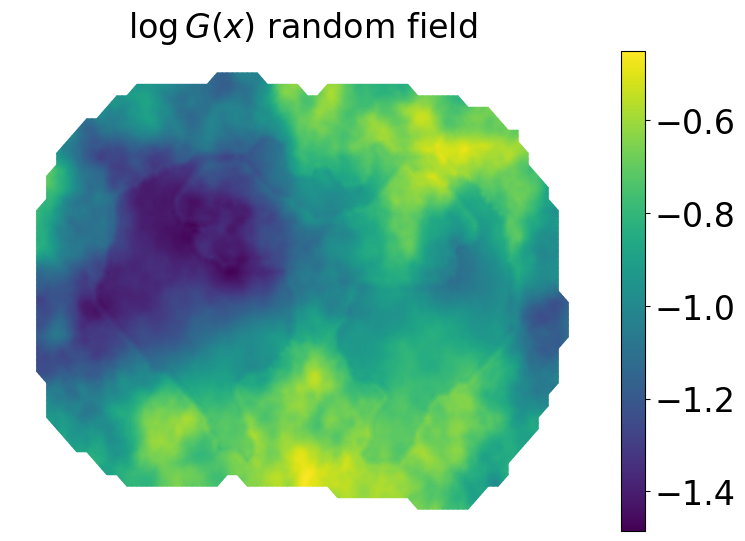}
   \caption{Left: Illustration of gray and white matter in a rat's brain. Middle: Mean of the prior distribution $m_\text{prior}$. Right: A random sample drawn from the prior distribution $m \sim \cN (m_\text{prior}, \cC_\text{prior})$.}
    \label{fig:3plots}
\end{figure}

The parameter $D$ characterizes tumor diffusion, encompassing invasion and cell migration processes, while $G$ represents the tumor's growth rate, capturing the proliferation of tumor cells through division and expansion. 
We use the parameters from \cite{liang2023bayesian} to define the prior distribution. The brain is divided into regions of gray and white matters, each with distinct characteristics, see the left part of Figure \ref{fig:3plots}. 

In our experiment, we take
$D$ as a constant in each region with $\log(D_\text{gm}) = -0.9937, \log(D_\text{wm}) = -0.3006$, and consider $G$ as a random field with log-normal distribution $\log(G) = m \sim \cN(m_{\text{prior}}, \cC_\text{prior})$ with the mean $m_{\text{prior}}$ given in Table \ref{table:params} and a Mat\'ern covariance operator $\cC_\text{prior} = (-\gamma \Delta + \delta I )^{-2}$ with $\gamma = \rho/(4\sqrt{2\pi}\sigma)$ and $\delta = \sqrt{2}/(\sigma \rho \sqrt{\pi})$, with the variance $\sigma^2$ and correlation length $\rho$ in the two regions reported in Table \ref{table:params}.

\begin{table}[!htb]
\centering
\caption{Estimated hyper-parameters of the tumor growth model.}
\begin{tabular}{cc|cc}
\hline
\multicolumn{4}{c}{Prior mean and variance of parameters} \\
\hline
\multicolumn{2}{c|}{$\log(G_\text{gm})$} & \multicolumn{2}{c}{$\log(G_\text{wm})$} \\
\multicolumn{2}{c|}{$\log(1/\text{day})$} & \multicolumn{2}{c}{$\log(1/\text{day})$} \\
Mean & Variance & Mean & Variance \\
\hline
-0.7800 & 0.0682 & -0.8419 & 0.0682 \\
\hline
\multicolumn{4}{c}{Spatial correlation lengths of $G$} \\
\hline
\multicolumn{2}{c|}{$\rho_\text{gm}$ (mm)} & \multicolumn{2}{c}{$\rho_\text{wm}$ (mm)} \\
\multicolumn{2}{c|}{6.0} & \multicolumn{2}{c}{12.0} \\
\hline
\end{tabular} \label{table:params}
\end{table}

\begin{figure}[!htb]
    \centering
    \includegraphics[width=0.3\textwidth]{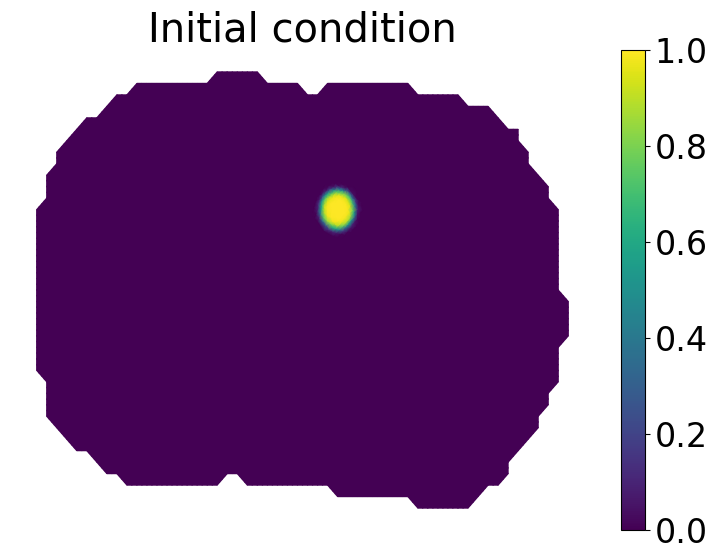}
    \hspace{0.02\textwidth}
    \includegraphics[width=0.3\textwidth]{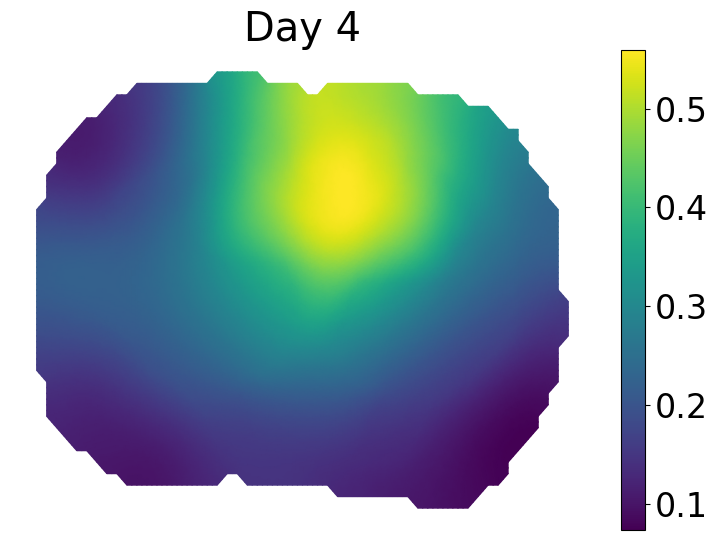}
    \hspace{0.02\textwidth}    
    \includegraphics[width=0.30\textwidth]{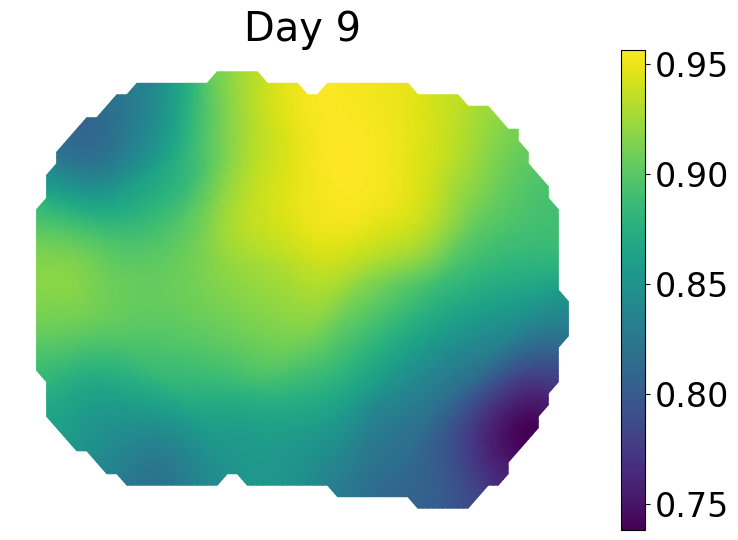}
   \caption{Left: Initial tumor implantation at day $t_0 = 0$. The volume fraction of the tumor at day $t_{40} = 4$ (middle) and day $t_{90} = 9$ (right) at a random sample of the parameter.}
    \label{fig:states}
\end{figure}

The initial condition represents the tumor implantation in the brain at $t=0$, as shown in the left part of Figure~\ref{fig:states}. We solve the PDE over $T = 10$ days using a FEM with piecewise linear 
finite element with $14,003$ degrees of freedom and an implicit time stepping with a uniform time step size of $\Delta t = 0.1$, which results in $K = 100$ time steps. Figure~\ref{fig:states} illustrates the volume fraction of the tumor at time $t_0 = 0$, $t_{40} = 4$, and $t_{90} = 9$, obtained as the solution of the PDE at a random sample of $m$. The SBOED problem is to select 4 out of 10 days, at time $t_{10}, t_{20}, \dots, t_{100}$, to take MRI images adaptively to infer the parameter $m$ accurately.

\subsection{Dimension reduction}

\begin{figure}[!htb]
    \centering
    \begin{tabular}{cccc}
        \includegraphics[width=0.24\textwidth]{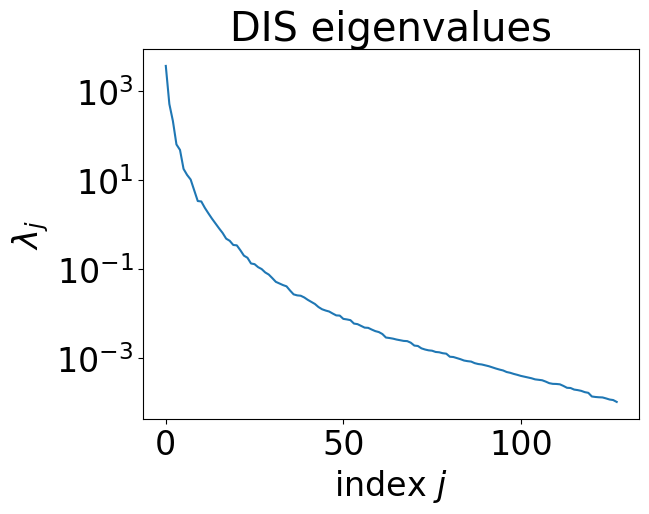} 
        \includegraphics[width=0.24\textwidth]{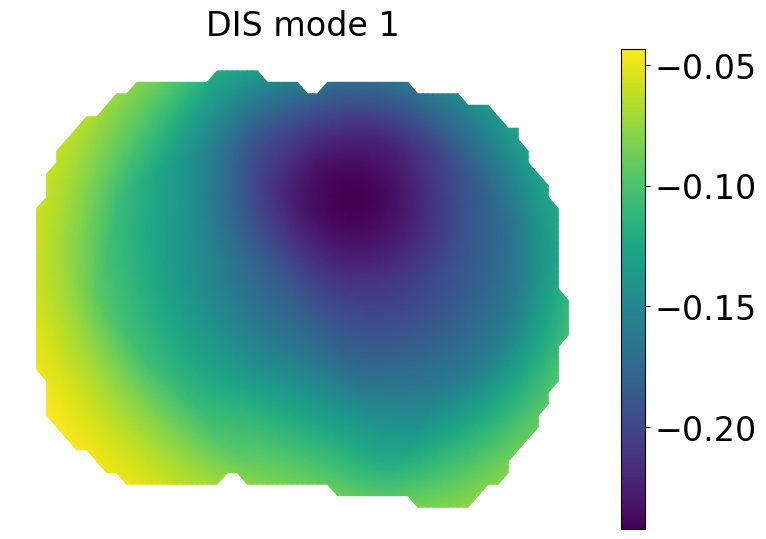} &
        \includegraphics[width=0.24\textwidth]{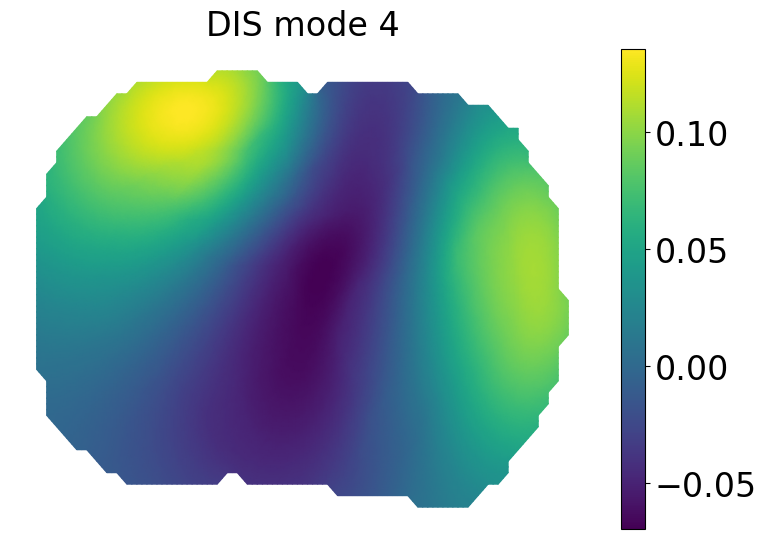} &
        \includegraphics[width=0.24\textwidth]{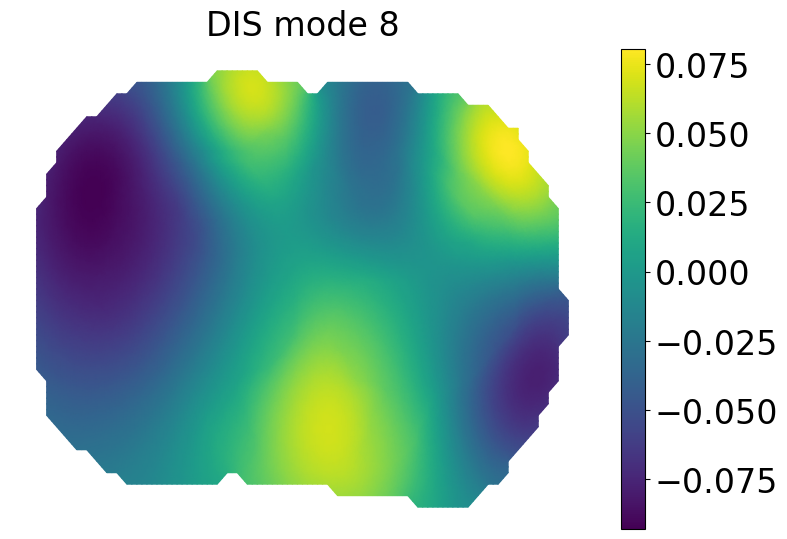}\\
        \includegraphics[width=0.24\textwidth]{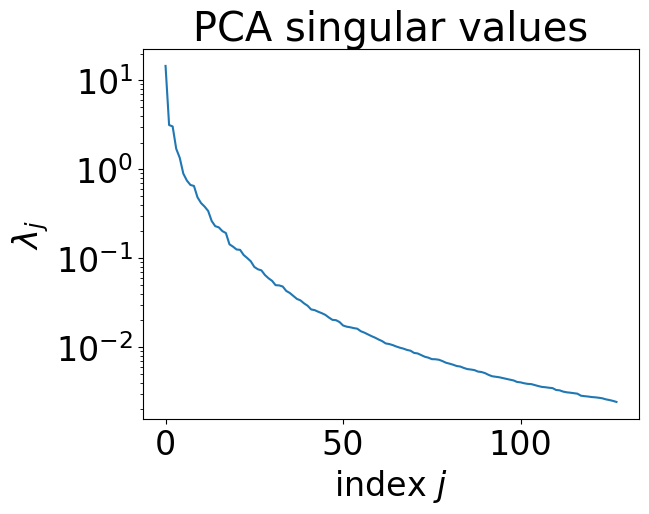} 
        \includegraphics[width=0.24\textwidth]{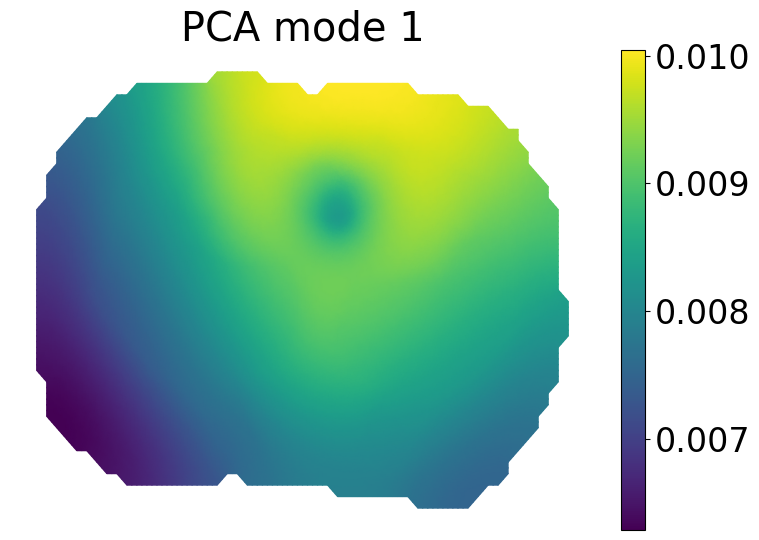} &
        \includegraphics[width=0.24\textwidth]{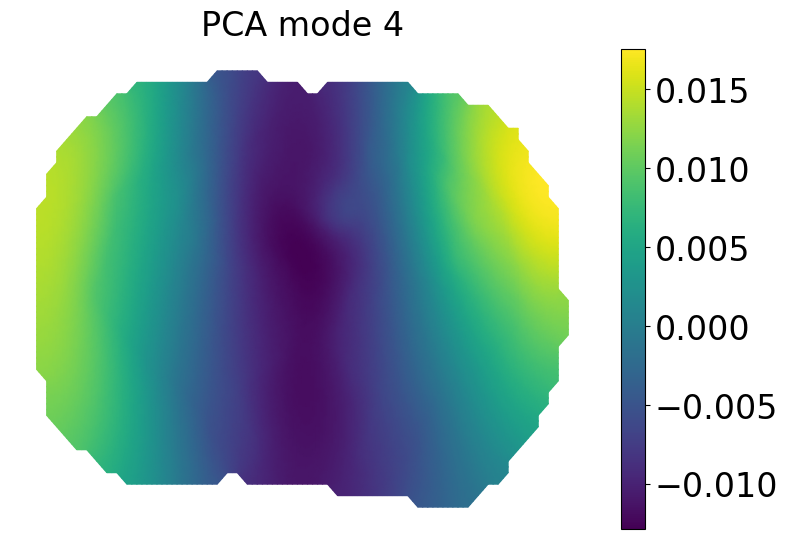} &
        \includegraphics[width=0.24\textwidth]{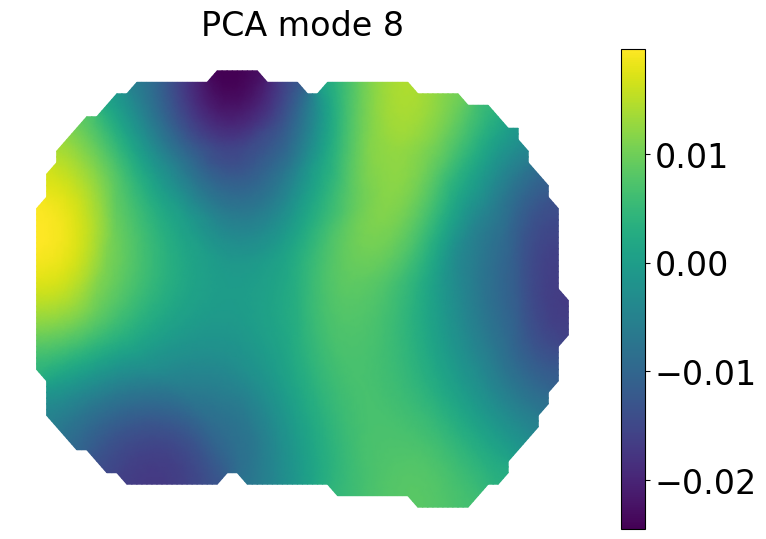}
    \end{tabular}
    \caption{Decay of the eigenvalues of DIS for input parameter dimension reduction (top left) and singular values (bottom left) by SVD for PCA output dimension reduction, and their corresponding modes. }
    \label{fig:modes}
\end{figure}

The dimension of both the discretized model parameter and the observation (we use the full state from the MRI) is $14,003,$ which is high. 
We perform the dimension reduction as in Section \ref{sec:dim-red}, where we use $256$ samples to compute the expectation in \eqref{eq:eigenias} for DIS and $1,024$ samples to generate the snapshot matrix $\mathbb{B}$ in \eqref{eq:PCA} for the PCA. The eigenvalues and modes of the DIS \eqref{eq:DIS} dimension reduction for the input parameter and the singular values and modes of the PCA \eqref{eq:PCA} dimension reduction for the output observation are shown in Figure~\ref{fig:modes}. 
We observe that the eigenvalues of DIS and the singular values of PCA decay rapidly. For simplicity, we truncate the modes at $r = 64$ for both the input and output projections, leading to less than 1\% dimension reduction errors in both projections.

\subsection{Neural network approximations}

To benchmark our proposed LANO surrogate for the approximation of the PtO map and its Jacobian, we compare it to two other neural network surrogate models. One is neural ODE \cite{chen2018neural}, which models the evolution of the dynamical system as an ODE system using neural networks in the latent space, e.g., 
\begin{equation}
    \beta_{F_{k+1}} \approx \cN^{\text{ODE}}_\bstheta (\beta_{F_k}, \beta_\bsm), \quad k = 0, \dots, K-1,
\end{equation}
where $\cN^{\text{ODE}}_{\bstheta}$ is a neural network parameterized by $\bstheta$. The other one is a 
DIPNet \cite{o2022derivative}, which learns a map directly from the projected DIS coefficients to the projected PCA coefficients at each time step, e.g., 
\begin{equation}
    \beta_{F_k} \approx \cN^{\text{DIP}}_{\bstheta_k}(\beta_\bsm), \quad k = 1, \dots, K,
\end{equation}
where the neural networks $\cN^{\text{DIP}}_{\bstheta_k}$,  parameterized by $\bstheta_k$ at each step $k$, 
are trained using both the PtO map and its Jacobian as in DINO \cite{o2024derivative}. 
We use three ResNet \cite{he2016deep} layers in both of these models, where the input and output dimensions are 64, and the ResNet layer width is taken as 100.

To evaluate the neural networks' performance, we consider two expected relative error metrics for the PtO map and the reduced Jacobian at each time step,
\begin{equation}\label{eq:errorFJ}
  \mathbb{E}_{\bsm} \left[ \frac{\|F_k(\bsm) - \Psi_F\cN_{k}(\beta_\bsm) - \bar{F} \|_\mathbb{M}}{\|\bsF_k(\bsm)\|_\mathbb{M}}\right] \text{ and } \mathbb{E}_{\bsm} \left[ \frac{\|\beta_{J_k} - \nabla_\beta{\cN}_{k}(\beta_\bsm)\|_F}{\|\beta_{J_k} \|_F}\right],
\end{equation}
where $\cN_k$ represents one of the three neural network models as a general notation, $\mathbb{M}$ denotes the mass matrix with $\|\bsy_k \|_\mathbb{M} = \sqrt{\bsy_k ^T \mathbb{M}\bsy_k}$, and $\|\cdot\|_F$ represents the Frobenius norm. Note that for the neural ODE, the reduced Jacobian can be computed recursively by the chain rule as  
\begin{equation}
    \nabla_\beta \cN_k(\beta_\bsm) = \frac{\partial \cN_\bstheta^{\text{ODE}}(\beta_{F_{k-1}}, \beta_\bsm)}{\partial \beta_\bsm} + \frac{\partial \cN_\bstheta^{\text{ODE}}(\beta_{F_{k-1}}, \beta_\bsm)}{\partial \beta_{F}} \nabla_\beta \cN_{k-1}(\beta_\bsm).
\end{equation}






After training the neural networks with $1,024$ training samples, we compute the relative errors for the PtO map and the reduced Jacobian with $100$ test samples for every tenth step, as reported in Table \ref{tab:comparison_methods}. 

\begin{table}[ht]
\centering
\resizebox{\textwidth}{!}{%
\begin{tabular}{|l|c|c|c|c|c|c|c|c|c|c|}
\hline
Day ($t_k$) / Step ($k$) & 1/10 & 2/20 & 3/30 & 4/40 & 5/50 & 6/60 & 7/70 & 8/80 & 9/90 & 10/100 \\
\hline
\multicolumn{11}{|c|}{Neural ODE} \\
\hline
One-step prediction (\%) & 3.24 & 3.24 & 2.68 & 2.33 & 2.02 & 1.90 & 1.86 & 1.82 & 1.74 & 1.63 \\
PtO map (\%) & 5.85 & 16.84 & 29.43 & 37.90 & 45.09 & 53.43 & 68.40 & 81.37 & 90.44 & 95.91 \\
Reduced Jacobian (\%) & 118.04 & 106.41 & 104.50 & 103.42 & 101.48 & 101.88 & 100.58 & 100.39 & 100.31 & 100.11 \\
\hline
\multicolumn{11}{|c|}{DIPNet} \\
\hline
PtO map (\%) & 13.26 & 30.70 & 48.62 & 66.50 & 80.07 & 89.65 & 92.36 & 86.60 & 69.29 & 60.00 \\
Reduced Jacobian (\%) & 8.15 & 5.99 & 6.30 & 7.58 & 12.22 & 22.74 & 39.12 & 56.94 & 79.81 & 89.06 \\
\hline
\multicolumn{11}{|c|}{LANO} \\
\hline
PtO map (\%) & 8.05 & 8.04 & 6.88 & 6.00 & 5.17 & 4.36 & 3.80 & 3.05 & 2.61 & 2.27 \\
Reduced Jacobian (\%) & 4.05 & 2.99 & 2.66 & 2.37 & 2.27 & 2.20 & 2.06 & 1.78 & 1.54 & 1.57 \\
\hline
\end{tabular}
}
\caption{Relative error (reported in \%) for the PtO map and the reduced Jacobian by neural ODE (top), DIPNet (middle), and our proposed method LANO (bottom) for 10 different time instances. Both DIPNet and LANO are trained with reduced Jacobian information.}
\label{tab:comparison_methods}
\end{table}

The neural ODE achieves high accuracy in one-step prediction for most iterations, with errors consistently below $4\%$. However, we observe significant error accumulation when applied recursively to predict the PtO map. The recursive prediction error starts at $5.85\%$ for the first $10$ iterations, but by the fourth day, this error escalates to $37.90\%$. 
This highlights a limitation of the neural ODE approach in maintaining accuracy over multiple recursive steps, posing a significant challenge for applications requiring long-term predictions or simulations. The DIPNet, on the other hand, shows varying performance across different time steps. On day 1, it achieves a PtO map error of 13.26\% and a reduced Jacobian error of $8.15\%$. As it does not build the nonlinear dynamical evolution in the architecture, the PtO map error quickly increases to $89.65\%$, and the reduced Jacobian error rises to $22.74\%$ on day 6.
In contrast, our proposed method LANO demonstrates superior stability and accuracy across time steps. The PtO map error peaks at $8.05\%$ on day 1, gradually decreasing to $2.27\%$ by day 10. The reduced Jacobian error shows less variation, ranging from $4.05\%$ to $1.57\%$ across all time steps. This comparison underscores the effectiveness of our approach in capturing the nonlinear dynamical evolution and achieving high accuracy over extended time horizons by the attention mechanism in using accumulative information from the dynamical process.

\begin{figure}[h]
    \centering
    \begin{tabular}{cccc}
        \includegraphics[width=0.24\textwidth]{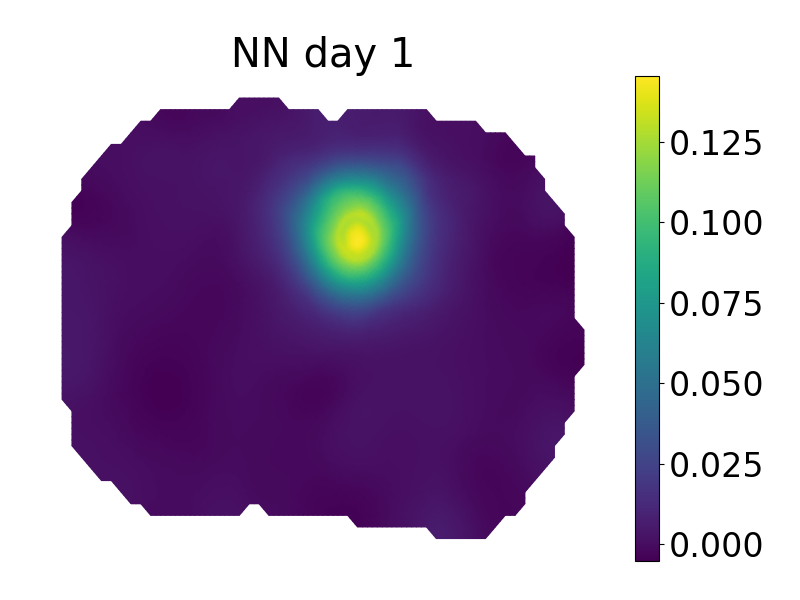} &
        \includegraphics[width=0.24\textwidth]{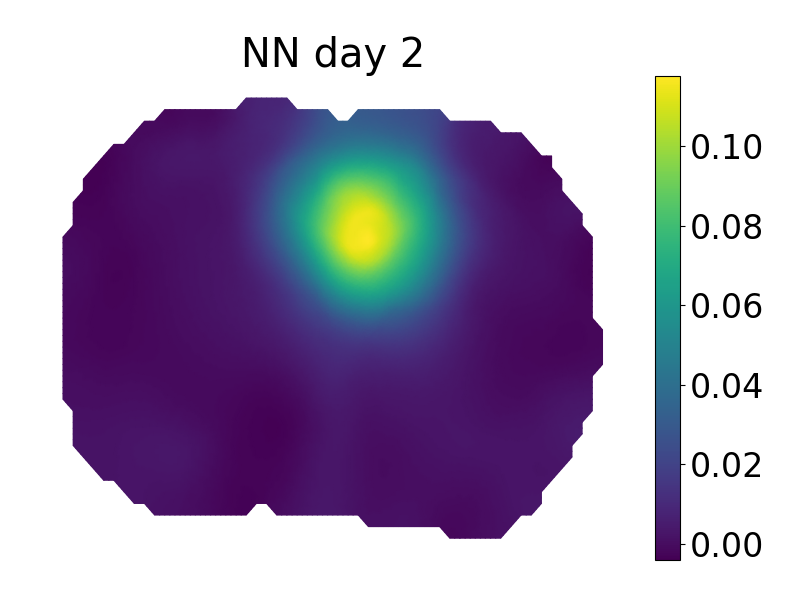} &
        \includegraphics[width=0.24\textwidth]{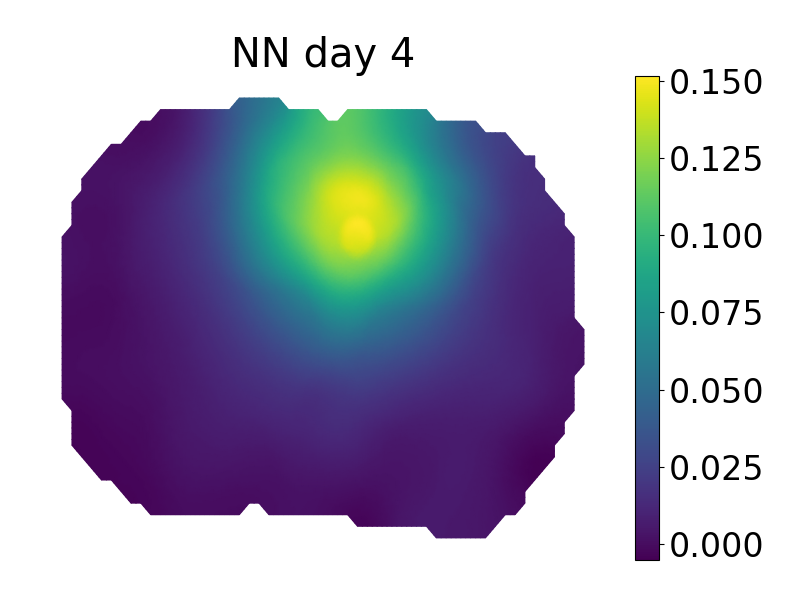} &
        \includegraphics[width=0.24\textwidth]{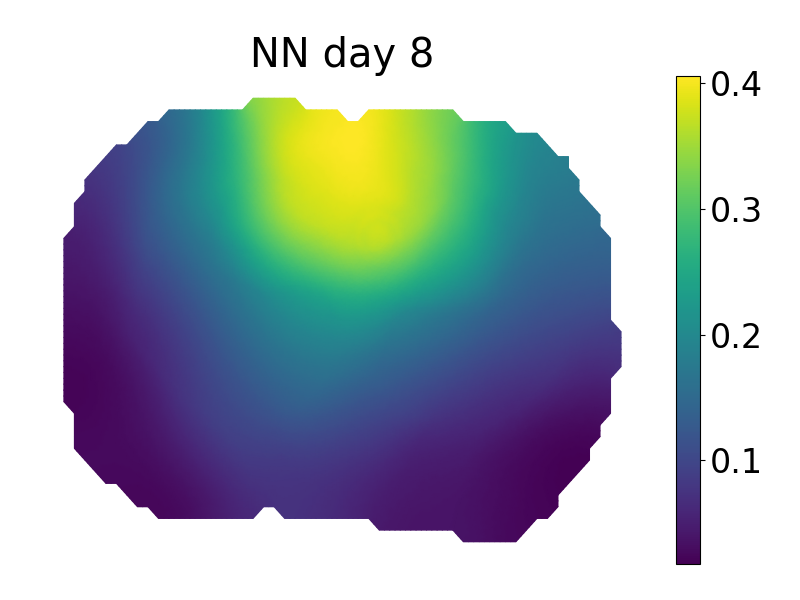} \\
        \includegraphics[width=0.24\textwidth]{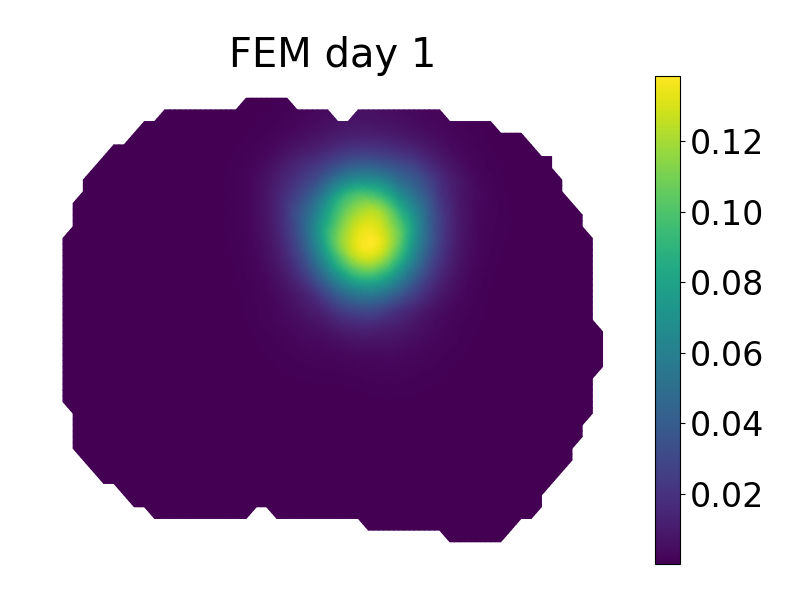} &
        \includegraphics[width=0.24\textwidth]{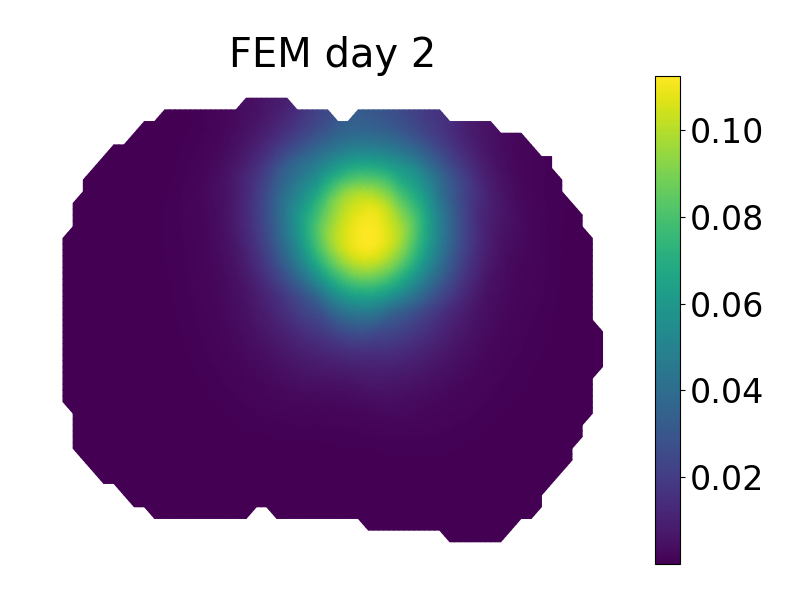} &
        \includegraphics[width=0.24\textwidth]{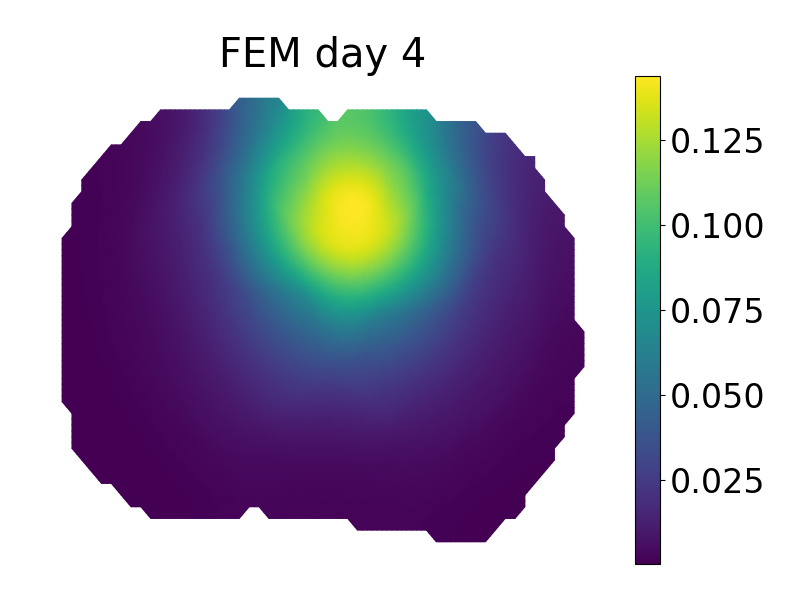} &
        \includegraphics[width=0.24\textwidth]{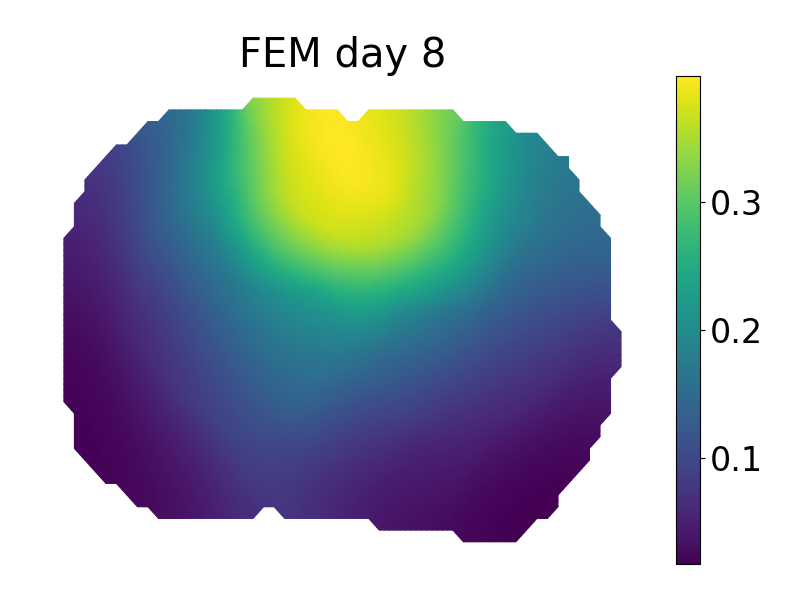} \\
        \includegraphics[width=0.24\textwidth]{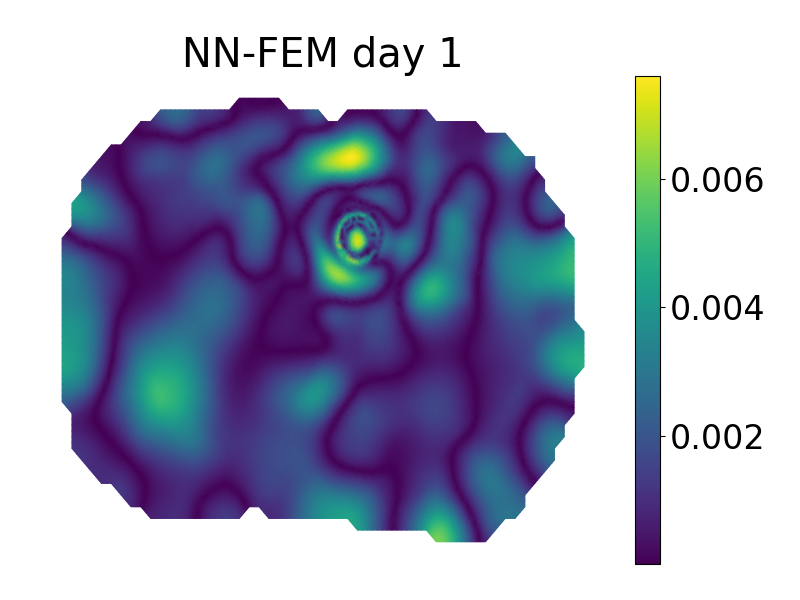} &
        \includegraphics[width=0.24\textwidth]{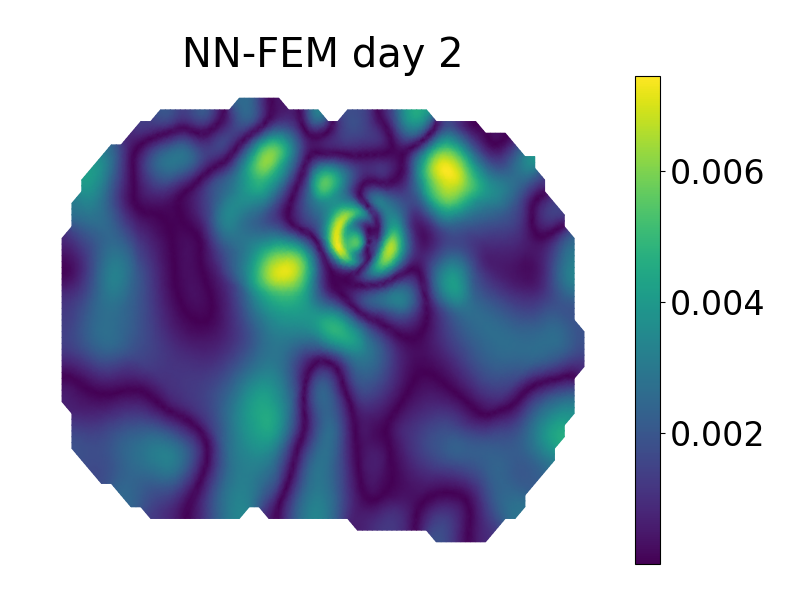} &
        \includegraphics[width=0.24\textwidth]{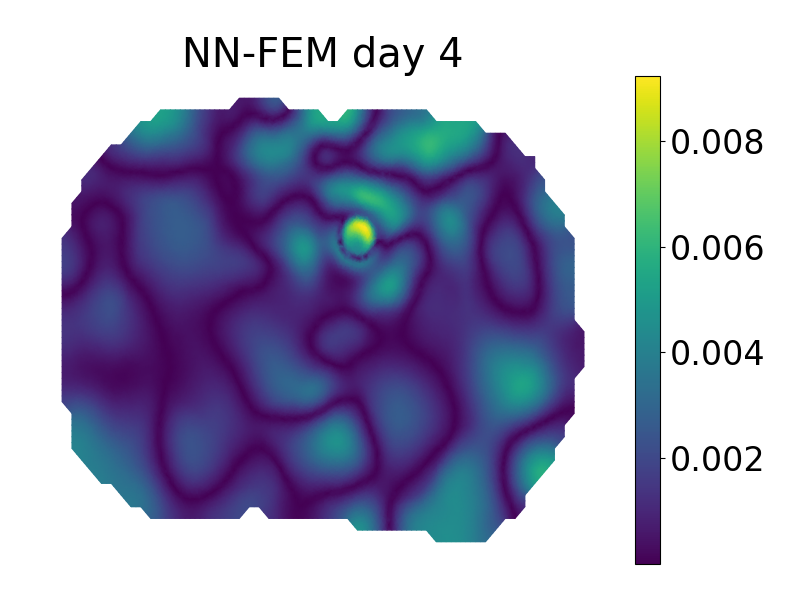} &
        \includegraphics[width=0.24\textwidth]{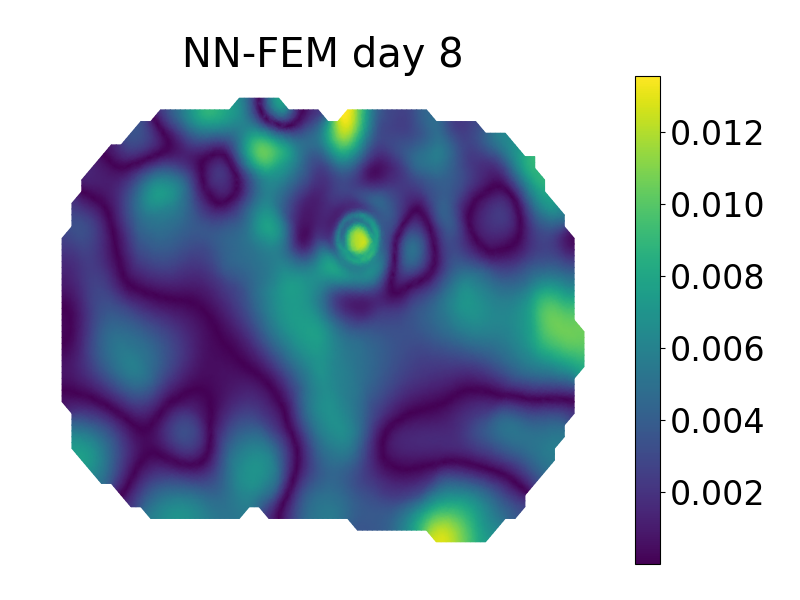} \\
    \end{tabular}
    \caption{Comparison of the approximation of the PtO map/state by our neural network (NN) surrogate and FEM computation. NN (top), FEM (middle), and their difference (bottom) on days 1, 2, 4, and 8.}
    \label{fig:nn_fem_comparison-diano}
\end{figure}

We also visualize the results at selected time steps [10, 20, 40, 80], which represent 1st, 2nd, 4th, and 8th day, in Figure~\ref{fig:nn_fem_comparison-diano}. This figure allows for a direct comparison between LANO approximations and FEM solutions, offering insights into how well the network captures the system's dynamics over time.

\subsection{Application to SBOED}
\begin{figure}[!htb]
    \centering
    \includegraphics[width=0.3\textwidth]{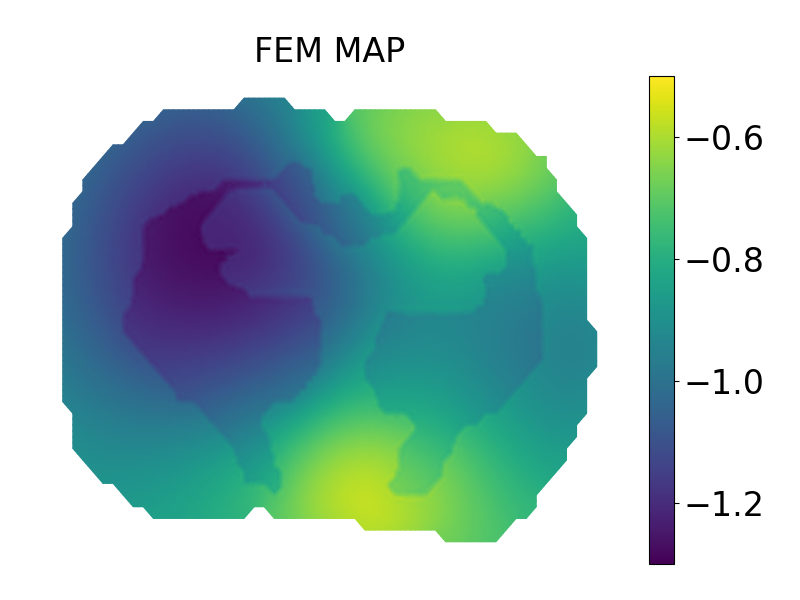} 
    \includegraphics[width=0.3\textwidth]{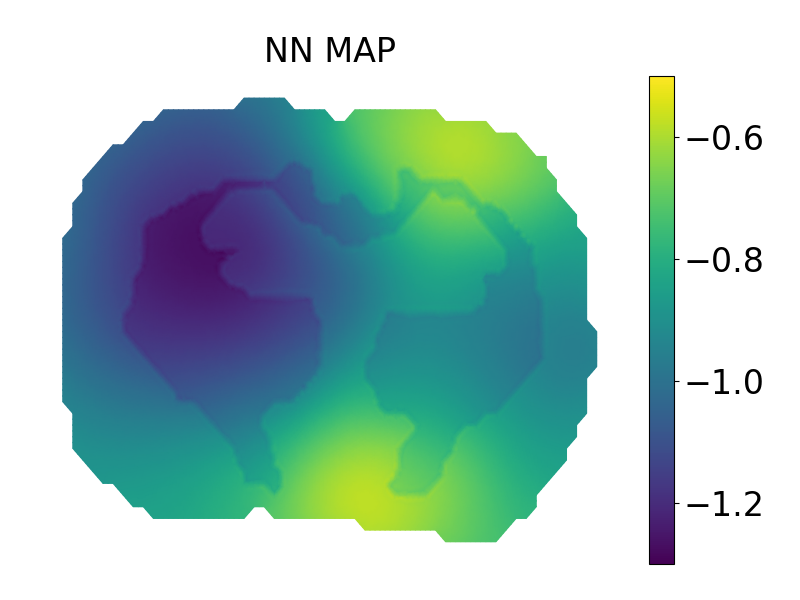} 
    \includegraphics[width=0.3\textwidth]
    {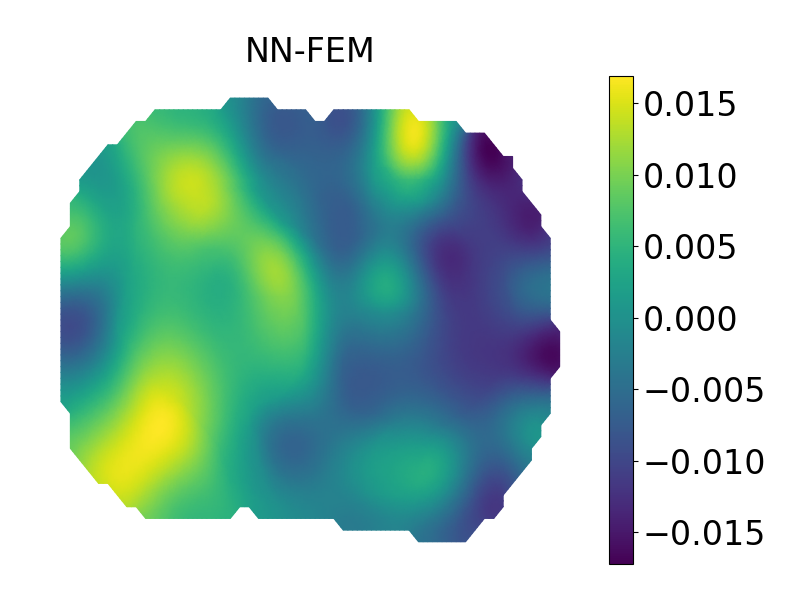} 

    \includegraphics[width=0.3\textwidth]{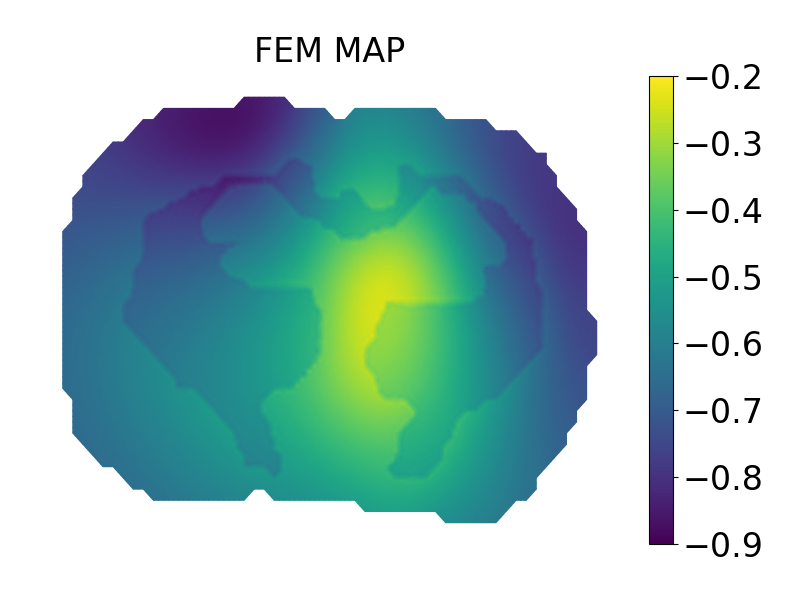} 
    \includegraphics[width=0.3\textwidth]{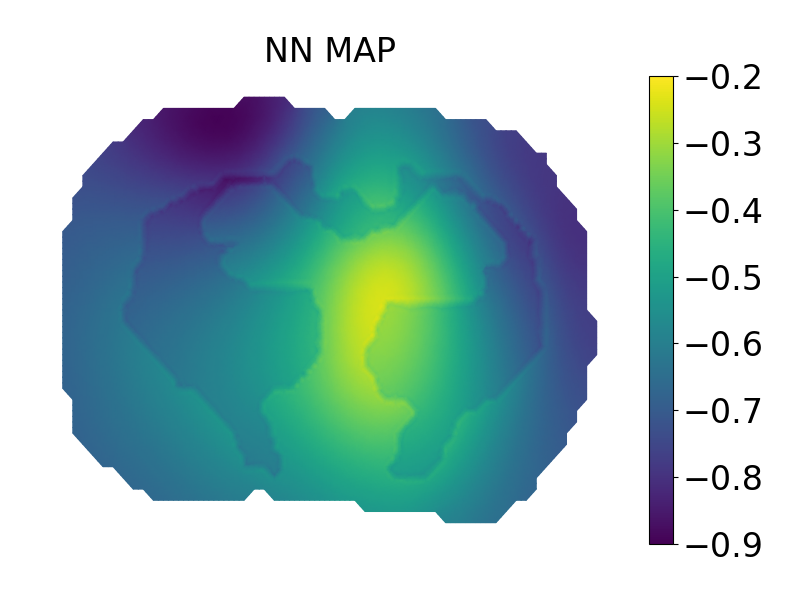} 
    \includegraphics[width=0.3\textwidth]{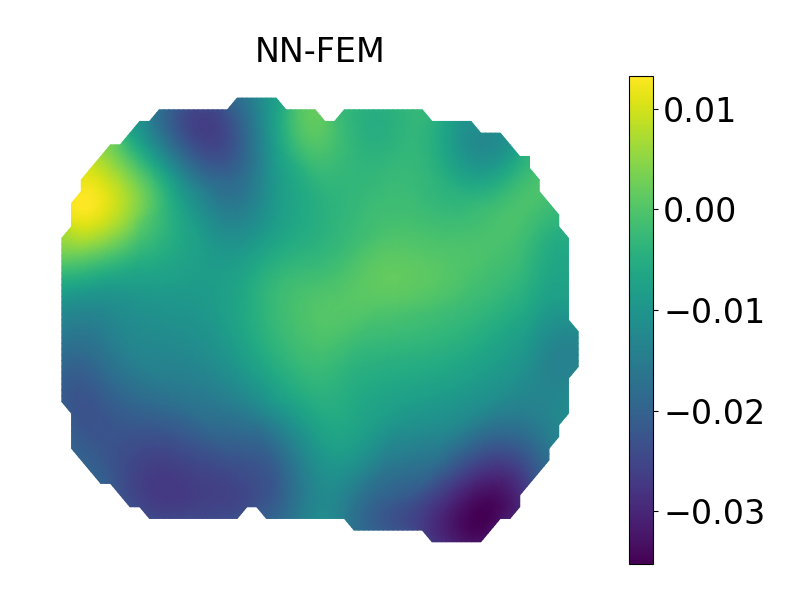}
    \caption{MAP points computed by FEM (left) and our NN surrogate (middle), and their difference (right) for a random sample drawn from the prior. Top: daily observations, bottom: observations at day 2, 5, 8.
    }
    \label{fig:MAP-compare-10}
\end{figure}

To further validate the effectiveness of our proposed neural network to solve the SBOED problem, we first employ it to compute the MAP point by solving the optimization problem \eqref{eq:inv-red} in the reduced space. To quantify the accuracy of the MAP point estimation, we define the relative error metric:
\begin{equation}\label{eq:MAP-re}
  \E_{\bsy} \left[ \frac{||\bsm_{\text{MAP}}^{\bsy, \xi} - \Psi_{\bsm}\beta_{\text{MAP}}^{\bsy, \xi} - \bsm_\text{prior}||_\mathbb{M}}{||\bsm_{\text{MAP}}^{\bsy, \xi}||_\mathbb{M}}\right],
\end{equation}
where $\bsm_{\text{MAP}}^{\bsy, \xi}$ represents the MAP point computed by FEM, $\Psi_{\bsm}\beta_{\text{MAP}}^{\bsy, \xi} + \bsm_\text{prior}$ is the neural network's approximation of the MAP point. To evaluate the expectation in \eqref{eq:MAP-re}, we generate $128$ random samples from the prior distribution, solve the dynamical system, generate the observation data $\bsy$ with Gaussian noise, and calculate $128$ MAP points from the observations collected once every day. Compared to FEM, our proposed method LANO achieves a mean relative error of $1.52\%$ and a standard deviation of $0.44\%$. See Figure~\ref{fig:MAP-compare-10} for the comparison of the MAP points computed by FEM and LANO at a random sample. 


To further evaluate the robustness and flexibility of our proposed surrogate, we also compute the MAP point using a subset of observations, e.g., $\{\bsy_{2}, \bsy_{5}, \bsy_{8}\}$. This sparse observation set allows us to assess the performance of our approach when dealing with limited data.
%
In this scenario, we achieve a mean relative error of 1.31\% with a standard deviation of 0.56\%. See Figure~\ref{fig:MAP-compare-10} for a comparison at one random sample. 

%
%

\begin{figure}[!htb]
    \centering
    \includegraphics[width=0.5\textwidth]{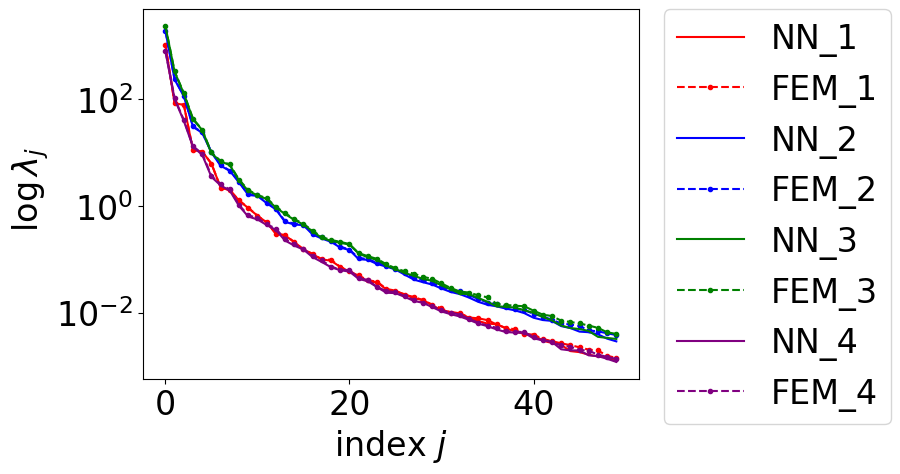}
    
   \caption{Comparison of the decay of the eigenvalues of \eqref{eq:geneig} computed by FEM and the eigenvalues of \eqref{eq:red-geigen} computed by the LANO NN surrogate at the same four observation data and random samples.}
    \label{fig:eigvals}
\end{figure}

To demonstrate the approximation accuracy of the eigenvalues used in the information gain \eqref{eq:reduced-klapprox_term}, we solve the generalized eigenvalue problem \eqref{eq:geneig} by FEM at the MAP points computed by FEM and solve the eigenvalue problem \eqref{eq:red-geigen} by LANO at the MAP points computed by LANO, where the MAP points are computed for the same observation data generated from the same prior samples. Figure \ref{fig:eigvals} displays the comparison of the decay of the eigenvalues for four random samples, which demonstrates very high accuracy of the eigenvalue approximation by the LANO surrogate. It achieves a mean relative error of $0.8\%$ with a standard deviation of $0.44\%$ over eight random samples in the evaluation of the first term of the information gain \eqref{eq:reduced-klapprox_term} that involves all the eigenvalues. In the evaluation of the second term that involves the MAP points, a mean relative error of $0.2\%$ and a standard deviation of $0.03\%$ are achieved by the LANO surrogate. These small errors collectively demonstrate the high accuracy in the approximation of the information gain.


\begin{figure}[!htb]
    \centering
        \includegraphics[width=0.24\textwidth]{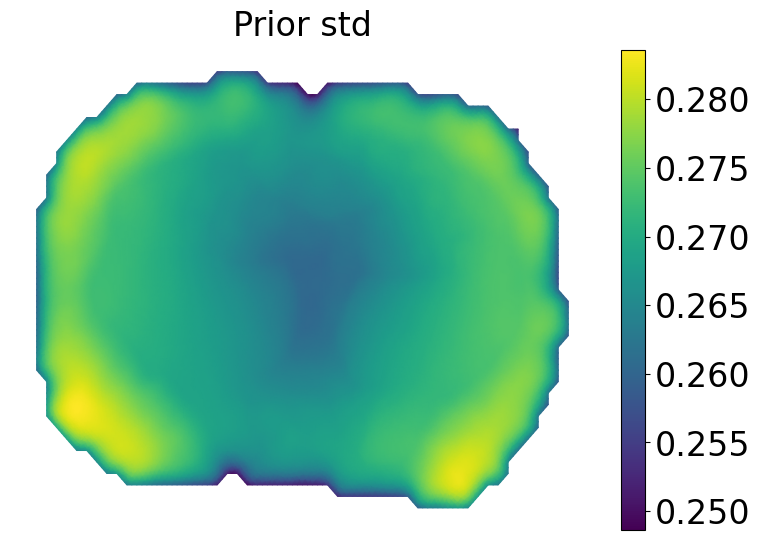}
        \includegraphics[width=0.24\textwidth]{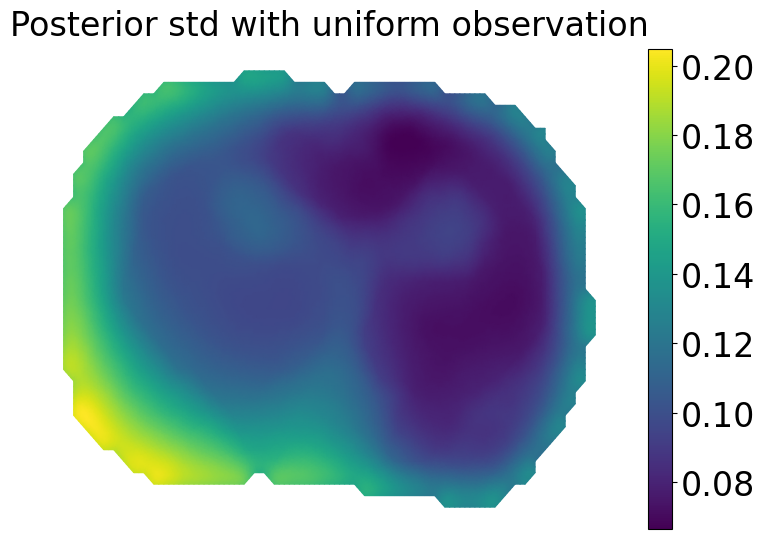}
        \includegraphics[width=0.24\textwidth]{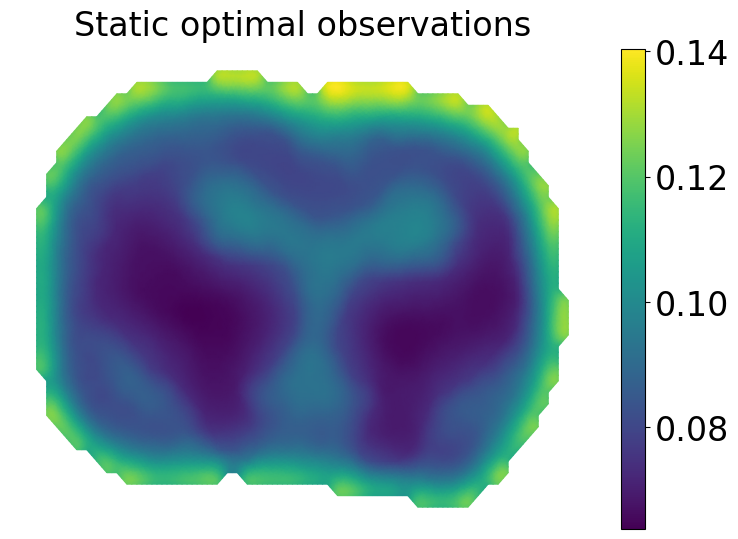}
        \includegraphics[width=0.24\textwidth]{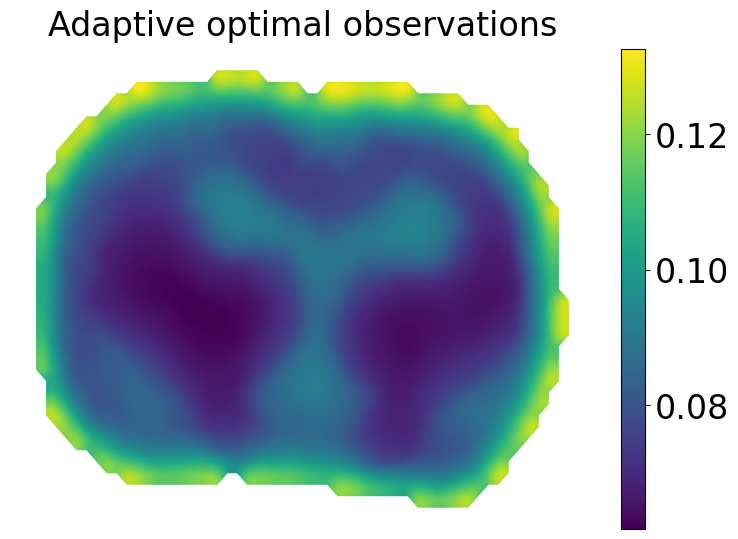}

    \caption{From left to right: pointwise standard deviation of the parameter field following the prior distribution and the posterior distributions with uniform design, static optimal design, and adaptive optimal design.}
    \label{fig:uq-decrease}
\end{figure}

With the highly accurate approximation of the information gain, we solve the adaptive SBOED problem \eqref{eq:terminalinitialBOED}. At $i = 1$, we obtain the optimal experimental design that make observations at day $2, 8, 9, 10$, which corresponds to a static SBOED \eqref{eq:staticBOED}. After the adaptive optimization with Algorithm \ref{alg:adaptive-incre}, the optimal experimental design changes to $2, 7, 9, 10$. The standard deviations of the parameter fields are shown in Figure \ref{fig:uq-decrease} corresponding to the prior, the posterior at an intuitive uniform design at day $2, 4, 6, 8$, the posterior at the static optimal design and adaptive optimal design. 
We observe that the adaptively optimized design results in most informative data with smallest uncertainty. 
The design with late stage observations implies that the tumor growth that spread over the domain at later stages is more informative for the parameter field in the entire domain.
In this example, the static SBOED and adaptive SBOED show a small difference as we use synthetic data from PDE simulation. We anticipate that adaptive SBOED would show greater advantages in scenarios with real-world observations that largely differ from the PDE simulation data.




\subsection{Efficiency of the computational framework}
\label{sec:efficiency}
In this section, we demonstrate the efficiency of the proposed method following the guidance in \cite{mcgreivy2024weak} to mitigate reporting biases. We measure not only the acceleration by LANO compared to FEM for the online evaluation but also report the offline computing time for the construction of LANO. 

Specifically, for the online evaluation cost, we report the time in computing the PtO map, the MAP point, the eigenpairs, and the information gain by FEM using the hIPPYlib package \cite{villa2021hippylib} and by LANO using PyTorch \cite{paszke2019pytorch}. 
We solve the state PDE with the following parameters: solver=SNES, absolute tolerance=1e-10, relative tolerance=1e-5, maximum iterations=100. To solve the optimization problem \eqref{eq:MAP} in computing the MAP point, we use the Newton-CG solver with relative tolerance=1e-4, absolute tolerance=1e-4, maximum iterations=100, and globalization=LS. 
To solve the generalized eigenvalue problem \eqref{eq:geneig}, we use a double pass randomized algorithm. When measuring time for LANO, 
we use L-BFGS optimizer for MAP point computation with the following parameters: maximum iterations=150, history size=150, tolerance grad=1e-7, tolerance change=1e-9 for MAP point computation. When computing the eigenpairs, we use the functorch automatic differentiation to compute Jacobian and construct the Gauss--Newton Hessian in the reduced space \eqref{eq:red-geigen}. We use AMD EPYC 7543 CPUs with 1 TB memory for FEM computation and LANO evaluation. For the neural network training, we use an NVIDIA RTX A6000 GPU with 48 GB memory.

Table \ref{tab:comparison} reports the comparison of the computational time and the corresponding speedup by LANO compared to FEM.
We observe that LANO achieves significant speed-ups in the computation of the PtO map ($388\times$) and especially the eigenpairs ($1364\times $), and a moderate spead-up of $57\times $ in the optimization for MAP point, and overall $197\times$ speed-up for the information gain. We remark that once the Laplace approximation of the posterior in Algorithm \ref{alg:compute-eig} is constructed by computing the MAP point and the eigenpairs, drawing each of the $N_s$ samples from the approximate posterior only takes 0.02 seconds by \eqref{eq:post-sample} in hIPPYlib and almost no time by \eqref{eq:reduced-post-sample} in PyTorch. 
\begin{table}[h]
    \centering
    \begin{tabular}{|l|c|c|c|c|}
        \hline
        time (s) &  PtO & MAP & Eigenpairs & Information Gain\\
        \hline
        FEM & 15.5 & 814.5 & 2,318.9 & 3,148.9 \\
        \hline
        LANO & 0.04 & 14.2 & 1.7 &  16.0 \\
        \hline \hline
        Speedup & 388$\times $ & 57$\times $ &  1364$\times $ & 197$\times $ \\
        \hline
    \end{tabular}
     \caption{Comparison of the time (in seconds) and the corresponding speedup by FEM and LANO for the computation of the PtO map, the MAP point, the eigenpairs, and the information gain. 
     }
    
    \label{tab:comparison}
\end{table}

We further report the offline time, which includes computing the bases, generating training data, and training the neural network. We summarize the time (in seconds) in Table \ref{tab:ct_nn_fem_2} with the parameters $N_t = 1,024$, $N_\bsm = 64$, $r_\bsm = r_F = 64$, $p = 10$. 
For the training of LANO, we use PyTorch AdamW optimizer with $0.001$ learning rate and $1,000$ epochs. The computational time is reported in Table \ref{tab:ct_nn_fem_2}. 
 
\begin{table}[!htb]
  \centering
  \begin{tabular}{|c|c|c|c||c|}
    \hline
    Bases & PtO & Jacobian & Total & Train (GPU) \\ \hline
      5,757 & 15,872 & 21,395 & 43,024 & 14,551 \\ \hline
  \end{tabular}
  \caption{Offline time (in seconds) in computing the input and output projection bases, PtO maps, and reduced Jacobians using AMD EPYC 7543, and training the neural network using NVIDIA RTX A6000.}
  \label{tab:ct_nn_fem_2}
\end{table}

The adaptive optimization by Algorithm \ref{alg:adaptive-incre} took $N_{\text{opt}} = 301$ 
evaluation of the conditional EIG, with each evaluation using $N_s = 128$ samples in Algorithm \ref{alg:compute-eig}. This would lead to an amortized computational speed up of $180\times$ by LANO compared to FEM in solving the SBOED problem, accounting for both the online evaluation and offline construction time. Note that this speed up is for the total computational time, not the wall clock time. In practice, we can parallelize the computation for both the offline construction and the online evaluation, e.g., we use 64 CPU processors in computing 1,024 training data. The amortized speed up for the wall clock time would depend on the number of available CPU processors and the  parallel algorithm. 




\section{Conclusion}\label{sec:conclusion}
In this work, we develop a new computational framework to solve infinite-dimensional SBOED problems constrained by large-scale PDE models. We propose an adaptive terminal formulation of the SBOED to achieve adaptive global optimality of the experimental design and establish an equivalent optimization problem with the EIG formulated as a conditional expectation of the KL divergence between the posterior at the terminal state and the prior at the initial state, which can be efficiently evaluated by low-rank Laplace approximation of the posteriors at both the terminal state and current state. 

We develop a derivative-informed LANO to approximate both the PtO maps and their Jacobians. LANO takes advantage of derivative-informed dimension reduction for latent encoding and an attention mechanism to capture the dynamics in the latent spaces of the parameter and observable. We formulate an efficient training of LANO using data from both the PtO maps and their Jacobians projected in the latent spaces. With a practical example of SBOED for tumor growth, we demonstrate the superior accuracy of our proposed method compared to two other surrogates for evaluating both the PtO maps and their Jacobians, which leads to its high accuracy in computing the MAP points 
and the eigenvalues in the evaluation of the optimality criteria. We also demonstrate the high efficiency of the proposed method that achieves an overall $180\times$ speed up in the total computational time for solving the SBOED problem, accounting for both the offline data generation and training time and the online evaluation time. 

In our SBOED problems, both Laplace approximation of the posterior and low-rank approximation of the posterior covariance significantly enhance computational efficiency. However, our method could be further developed for cases where the Laplace approximation is inadequate or the posterior covariance isn't low rank. We anticipate that techniques such as variational inference could accelerate the evaluation of optimality criteria in these challenging scenarios.
Additionally, our method has the potential for extension to more complex SBOED problems. These could involve selecting not only optimal observation times but also ideal spatial locations for sensor placement. For such extended problems, more efficient optimization strategies like greedy and swapping greedy algorithms \cite{wu2023fast,go2023accelerating} could be employed. Furthermore, future research is interesting on adaptively refining the neural network approximation to make predictions beyond the training horizon to enable a predictive digital twin of the physical system.

\section*{Acknowledgments}
We are grateful for the helpful discussions with Dr.\ Tobin Isaac, Dr.\ Xun Huan, and Grant Bruer.
This work is partially funded by National Science Foundation under grants \#2245674, \#2325631, \#2245111, and \#2233032. 

\appendix 

\section{Proof of Theorem \ref{thm:equivalence}}
\label{sec:equivalence}

\begin{proof}
Note that in the first step at $i = 1$, when there is no data being observed, the SBOED with the cumulative formulation \eqref{eq:cumulativeBOED} becomes a static SBOED in \eqref{eq:staticBOED}, which is equivalent to the SBOED with terminal formulation \eqref{eq:terminalinitialBOED}, see the proof in, e.g., \cite{shen2023variational, foster2020unified, ivanova2021implicit}. By the same argument, for any $i > 1$, let $\mu(m|\bsy^*_{1:i-1}, \bsxi_{1:i-1}^*)$ denote the posterior distribution of the model parameter conditioned on the observed data $\bsy_{1:i-1}^*$ from the optimized experimental design $\bsxi_{1:i-1}^*$ before time $t_i$, we have that the SBOED problem \eqref{eq:cumulativeBOED} in the cumulative formulation is equivalent to the following optimization problem
\begin{equation}\label{eq:terminalBOED}
    \bsxi_{i:K}^* = \arg\max_{\bsxi_{i:K}} \E_{\pi(\bsy_{i:K}|\bsxi_{1:i:K},\bsy_{1:i-1}^*)} 
    \left[
    \dkl(\mu(m|\bsy_{1:i:K}, \bsxi_{1:i:K})||\mu(m|\bsy^*_{1:i-1}, \bsxi_{1:i-1}^*))\right], \quad i = 1, \dots, K,
\end{equation}
where $\mu(m|\bsy^*_{1:i-1}, \bsxi_{1:i-1}^*)$ is taken as the initial distribution at time $t_i$. 
In this following, we aim to show that the objective function in \eqref{eq:terminalBOED} is only different from that in \eqref{eq:terminalinitialBOED} by a constant $\dkl(\mu(m|\bsy_{1:i-1}^*,\bsxi_{1:i-1}^*)||\mu(m))$, so the two optimization problems are equivalent. By definition of the KL divergence, we have
\begin{equation}
\begin{split}
    & \E_{\pi(\bsy_{i:K}|\bsxi_{1:i:K},\bsy_{1:i-1}^*)}\left[\dkl(\mu(m|\bsy_{1:i:K},\bsxi_{1:i:K})||\mu(m|\bsy_{1:i-1}^*,\bsxi_{1:i-1}^*))\right]\\
    &= \E_{\pi(\bsy_{i:K}|\bsxi_{1:i:K},\bsy_{1:i-1}^*)}\left[\int_{M}\log\left(\frac{d\mu(m|\bsy_{1:i:K},\bsxi_{1:i:K})}{d\mu(m|\bsy_{1:i-1}^*,\bsxi_{1:i-1}^*)}\right)d\mu(m|\bsy_{1:i:K},\bsxi_{1:i:K})\right]\\
    &= \E_{\pi(\bsy_{i:K}|\bsxi_{1:i:K},\bsy_{1:i-1}^*)}\left[\int_{M}\log\left(\frac{d\mu(m|\bsy_{1:i:K},\bsxi_{1:i:K})}{d\mu(m)}\right)d\mu(m|\bsy_{1:i:K},\bsxi_{1:i:K})\right] \\
    & - \E_{\pi(\bsy_{i:K}|\bsxi_{1:i:K},\bsy_{1:i-1}^*)}\left[\int_{M}\log\left(\frac{d\mu(m|\bsy_{1:i-1}^*,\bsxi_{1:i-1}^*)}{d\mu(m)}\right)d\mu(m|\bsy_{1:i:K},\bsxi_{1:i:K})\right],
\end{split}
\end{equation}
where we multiplied and divided $d\mu(m)$ for the second equality. Note that the first term is the same as the objective function \eqref{eq:terminalinitialBOED}. 
For the second term, we have
\begin{equation}
\begin{split}
    & \E_{\pi(\bsy_{i:K}|\bsxi_{1:i:K},\bsy_{1:i-1}^*)}\left[\int_M \log\left(\frac{d\mu(m|\bsy_{1:i-1}^*,\bsxi_{1:i-1}^*)}{d\mu(m)}\right)d\mu(m|\bsy_{1:i:K},\bsxi_{1:i:K})\right] \\[4pt]
    & = \int_{\mathcal{Y}_{i:K}} \left(\int_M \log\left(\frac{d\mu(m|\bsy_{1:i-1}^*,\bsxi_{1:i-1}^*)}{d\mu(m)}\right)d\mu(m|\bsy_{1:i:K},\bsxi_{1:i:K})\right) \pi(\bsy_{i:K}|\bsxi_{1:i:K},\bsy_{1:i-1}^*) d\bsy_{i:K}\\[4pt]
    &= \int_{M}\left(\int_{\mathcal{Y}_{i:K}}{\pi( \bsy_{i:K}|m, \bsxi_{1:i:K},\bsy_{1:i-1}^*)}d\bsy_{i:K} \right) \log\left(\frac{d\mu(m|\bsy_{1:i-1}^*,\bsxi_{1:i-1}^*)}{d\mu(m)}\right)d\mu(m|\bsy_{1:i-1}^*,\bsxi_{1:i-1}^*) \\[4pt]
    &= \int_{M} \log\left(\frac{d\mu(m|\bsy_{1:i-1}^*,\bsxi_{1:i-1}^*)}{d\mu(m)}\right)d\mu(m|\bsy_{1:i-1}^*,\bsxi_{1:i-1}^*) \\[4pt]
    &= \dkl(\mu(m|\bsy_{1:i-1}^*,\bsxi_{1:i-1}^*)||\mu(m)), 
\end{split}
\end{equation}
with $\mathcal{Y}_{i:K} = (\mathcal{Y}, \dots, \mathcal{Y})$ and $\mathcal{Y} = \mathbb{R}^{d_y}$, where for the second equality we used the Bayes' rule 
\begin{equation}
\frac{d\mu(m|\bsy_{1:i:K},\bsxi_{1:i:K})}{d\mu(m|\bsy_{1:i-1}^*,\bsxi_{1:i-1}^*)} = \frac{\pi( \bsy_{i:K}|m, \bsxi_{1:i:K},\bsy_{1:i-1}^*)}{\pi(\bsy_{i:K}|\bsxi_{1:i:K},\bsy_{1:i-1}^*)}
\end{equation}
with the likelihood in the numerator and the marginal likelihood in the denominator, and for the third equality we used that the likelihood function is a probability density function in the data, which integrates to 1. This concludes the proof by noting that $\dkl(\mu(m|\bsy_{1:i-1}^*,\bsxi_{1:i-1}^*)||\mu(m))$ is a constant.  


\end{proof}

\section{Computation of derivatives}
\label{ap:Jac}

At time $t$, let $\nabla_m u(t): M \to V$ denote the Jacobian of the solution $u$ of \eqref{eq:pde} with respect to the parameter $m$, and let $(\nabla_m u(t))^T: V \to M$ denote its transpose, which satisfies
\begin{equation}
    ((\nabla_m u(t))^T \; v, m)_M = (v, \nabla_m u(t) \; m)_V , \quad \forall m \in M, \; \forall v \in V,
\end{equation}
where $(\cdot, \cdot)_V$ and $(\cdot, \cdot)_M$ are the inner products in Hilbert spaces $V$ and $M$. 
The goal is to compute the following quantity at any time $t \in [0, T]$, 
\begin{equation}\label{eq:JTJm}
    (\nabla_m u(t))^T \nabla_m u(t) \; \hat{m}, \quad \forall \hat{m} \in M.
\end{equation}

By taking derivative of \eqref{eq:pde} with respect to $m$, we obtain 
\begin{equation}
    \partial_t \nabla_m u + \partial_u R(u, m) \; \nabla_m u + \partial_m R(u, m) = 0,
\end{equation}
where we assume that $\partial_t \nabla_m u$ and $\nabla_m \partial_t u $ are continuous in $t$ and $m$ to have the change of order $\nabla_m \partial_t u = \partial_t \nabla_m u$, and $R$ is differentiable in $u$ and $m$ with the linear derivative operators $\partial_u R: V \to V'$ and $\partial_m R: M \to V'$. Under the assumption that the operator $\partial_t + \partial_u R(u, m)$ is invertible with its inverse map $(\partial_t + \partial_u R(u, m))^{-1}: V' \to V$, we obtain 
\begin{equation}\label{eq:J}
    \nabla_m u = -(\partial_t + \partial_u R(u, m))^{-1} \partial_m R(u, m)
\end{equation}
and its transpose as 
\begin{equation}\label{eq:JT}
    (\nabla_m u)^T = -(\partial_m R(u,m))^* (\partial_t + \partial_u R(u, m))^{-*},
\end{equation}
where the adjoint operator $(\partial_m R)^*: V\to M'$ satisfies 
\begin{equation}\label{eq:dRdmT}
    \langle(\partial_m R)^* \; v, m\rangle_{M' \times M} = \langle v, \partial_m R \; m\rangle_{V \times V'}, \quad \forall m \in M, \; \forall v \in V,
\end{equation}
with $\langle \cdot, \cdot \rangle_{M' \times M}$ and $\langle \cdot, \cdot \rangle_{V \times V'}$ denoting the duality pairings, and $(\partial_t + \partial_u R)^{-*}: V' \to V$, the inverse of the adjoint operator $(\partial_t + \partial_u R)^*: V \to V'$ that satisfies 
\begin{equation}\label{eq:adjointR}
    \langle(\partial_t + \partial_u R)^{*}\; w, v \rangle_{V' \times V} = \langle w, (\partial_t + \partial_u R) \; v\rangle_{V \times V'}, \quad \forall v, w\in V.
\end{equation}

To compute $(\nabla_m u(t))^T \nabla_m u(t) \; \hat{m}$ in \eqref{eq:JTJm} for any time $t \in [0, T]$ and parameter $\hat{m} \in M$, we first note that $\nabla_m u(t)$ vanishes at $t = 0$ because of the independence of initial condition $u_0$ on $m$. Then we compute $(\nabla_m u(t))^T \nabla_m u(t) \; \hat{m}$ by first computing $\hat{u}(t):=\nabla_m u(t) \; \hat{m} \in V$ with zero initial condition $\hat{u}(0) = 0$, and then computing $(\nabla_m u(t))^T \hat{u}(t)$. To do so, we need to solve the equation \eqref{eq:pde}, which can be discretized in time by, e.g., a backward Euler scheme with $t_k = \Delta t k $, $k = 0, \dots, K$ with $K = T/\Delta t$ (note that $K$ here could be much larger than the candidate observation times in the SBOED problem),
\begin{equation}\label{eq:state-t}
    \frac{u_{k+1} - u_k}{\Delta t}  + R(u_{k+1}, m) = 0, 
\end{equation}
with $k = 0, \dots, K-1$, and the notation $u_k = u(t_k)$ and a given initial condition $u_0$. This equation can be solved by a FEM in space $V_h \subset V$ with a Newton algorithm for nonlinear $R$ with respect to $u$. 

To compute $\hat{u}(t) = \nabla_m u(t) \; \hat{m}$ with $\nabla_m u(t)$ defined in \eqref{eq:J}, i.e., 
\begin{equation}\label{eq:hatu}
    (\partial_t + \partial_u R(u, m)) \; \hat{u} = -\partial_m R(u,m) \; \hat{m},
\end{equation}
we use the same time discretization and solve 
\begin{equation}
    \frac{\hat{u}_{k+1} - \hat{u}_k}{\Delta t} + \partial_u R(u_{k+1}, m) \; \hat{u}_{k+1} = -\partial_m R(u_{k+1}, m) \; \hat{m},
\end{equation}
with $k = 0, \dots, K-1$, and the initial condition $\hat{u}_0 = 0$,
which can be solved by a FEM in the same finite element space $V_h$ as for $u_{k+1}$. Then to compute $(\nabla_m u(t))^T \hat{u}(t)$ with $(\nabla_m u(t))^T$ defined in \eqref{eq:JT}, we first solve for $\hat{p}(t):=-(\partial_t + \partial_u R)^{-*} \hat{u}(t)$, which is equivalent to solve 
\begin{equation}\label{eq:phat}
    (\partial_t + \partial_u R(u, m))^* \; \hat{p}(t) = -\hat{u}(t).
\end{equation}
By the same time discretization, we obtain 
\begin{equation}\label{eq:qt}
    - \frac{\hat{p}_{k+1}-\hat{p}_k}{\Delta t} + (\partial_u R(u_k,m))^*\;\hat{p}_k = - \hat{u}_k, 
\end{equation}
with $k = K-1, \dots, 0$ going backward in time, and the terminal condition $\hat{p}_K = 0$,
which can be solved by a FEM in the same space $V_h$. Note that the adjoint operator $\partial_t^* = - \partial_t$ using integration by part in time.

Once $\hat{p}(t)$ is computed, we can apply $(\partial_m R(u, m))^*$ to $\hat{p}$ as required in \eqref{eq:JT}, see the relation \eqref{eq:dRdmT}, to conclude the computation as 
\begin{equation}
    (\nabla_m u(t))^T \nabla_m u(t) \; \hat{m} = (\partial_m R(u,m))^* \hat{p}.
\end{equation}

For a parameter-to-observable map $\mathcal{F}_t(m) = \mathcal{B} u(t)$ with $u(t)$ implicitly depending on $m$ through the equation \eqref{eq:pde}, we have
\begin{equation}
    (\partial_m \mathcal{F}_t(m))^T \partial_m \mathcal{F}_t(m) \; \hat{m} =  (\nabla_m u(t))^T \mathcal{B}^T \mathcal{B} \nabla_m u(t) \; \hat{m},
\end{equation}
for which we can follow the same computation as for $(\nabla_m u)^T \nabla_m u \; \hat{m}$, except that we need to apply the operator $\mathcal{B}^T \mathcal{B}$ to $\hat{u} = \nabla_m u \; \hat{m}$ before applying $(\nabla_m u)^T$, which changes \eqref{eq:qt} as 
\begin{equation}\label{eq:qtbb}
    - \frac{\hat{p}_{k+1}-\hat{p}_k}{\Delta t} + (\partial_u R(u_k,m))^*\;\hat{p}_k = - \mathcal{B}^T \mathcal{B} \; \hat{u}_k, 
\end{equation}
with $k = K-1, \dots, 0$ going backward in time, and the terminal condition $\hat{p}_K = 0$. Given $\hat{p}(t)$, we can compute 
\begin{equation}\label{eq:JtJmhat}
    (\partial_m \mathcal{F}_t(m))^T \partial_m \mathcal{F}_t(m) \; \hat{m} =  (\partial_m R(u,m))^* \hat{p}(t).
\end{equation}

	\bibliographystyle{apalike}
	\bibliography{bayesiancitations1,peng}

\end{document}